\documentclass[trackchanges]{aastex701}

\begin{document}

\title{Kinematics and Untwisting Motion of an Intriguing Jet-like Prominence Eruption}

\author[orcid=0000-0002-1509-3970,gname=Pradeep, sname='Kayshap']{Pradeep Kayshap} 
\affiliation{School of Advanced Sciences and Languages, VIT Bhopal University, Kothrikalan, Sehore, Madhya Pradesh - 466114, India}
\email[show]{virat.com@gmail.com}

\author[gname=Petr,sname=Jel\'\i nek]{Petr Jel\'\i nek}
\affiliation{Institute of Physics, University of South Bohemia, Ceske Budejovice, Czech Republic}
\email{pjelinek@prf.jcu.cz }

\author[0009-0007-1911-797X]{B. Suresh Babu}
\affiliation{School of Advanced Sciences and Languages, VIT Bhopal University, Kothrikalan, Sehore, Madhya Pradesh - 466114, India}
\email{sureshsuju96@gmail.com }

\author[sname='Baral']{Ashok Kumar Baral}
\affiliation{School of Advanced Sciences and Languages, VIT Bhopal University, Kothrikalan, Sehore, Madhya Pradesh - 466114, India}
\email{ashok.kumar@vitbhopal.ac.in}

\author[orcid=0000-0001-9493-4418,gname=Yuandeng, sname='Shen']{Yuandeng Shen}
\affiliation{State Key Laboratory of Solar Activity and Space Weather, School of Aerospace, Harbin Institute of Technology, Shenzhen 518055, People's Republic of China}
\email[show]{ydshen@hit.edu.cn}



\begin{abstract}
We aim to investigate the blowout jet-like prominence eruption, which occurred on October 6$^{th}$, 2023, with the help of imaging and spectroscopic observations. Firstly, the prominence rises slowly with a speed of 33 km/s, followed by a fast rise (i.e., 338 km/s). Later, the northern leg breaks completely, and the eruption forms the blowout jet. The jet consists of different plasma threads, which show a range of upflow (i.e., 125 to 593 km/s) and downflow velocities (i.e., 43 to 158 km/s). The jet plasma column exhibits transverse oscillations, and this motion (untwisting motion) propagate at the speed of 267 km/s, are 
consistent with being Alfev{\'e}n waves. The transverse motion has the time period, amplitude, and transverse velocity of 1332 s, 26.19 Mm, and 126.18$\pm$7.27 km/s, respectively, and this transverse oscillation decays over time. Interestingly, the different plasma threads within the jet's body exhibit decayless transverse oscillations, and these decayless oscillations are related to the main decaying transverse oscillation. The transverse velocity of these decayless oscillations ranges from 66 to 30 km/s, the amplitudes from 8.52 to 2.74 Mm, and periods from 811 to 406 s. In addition, the spectroscopic analysis reveals Si~{\sc iv} lines are forming in the optically thick conditions in high electron density regions (i.e., near the base of the blowout jet). Lastly, we mention that two weak C-class flares occurred during this event, and further, one CME also occurred, which propagated with the speed of $\sim$250 km/s. 
\end{abstract}


\section{Introduction} 
The prominences, which are made of cool material, appear out of the limb, and they are studied in great detail; see the review articles by \cite{1995ASSL..199.....T}, \cite{2010SSRv..151..243L}, \cite{2010ApJ...720..757M}, \cite{2014LRSP...11....1P}, \cite{2018LRSP...15....7G}, and \cite{2023SSRv..219...33G}. In high spatial resolution, the prominences show more complex fine structures and dynamics that need to be understood further (e.g., \citealt{2015ApJ...814L..17S, 2018ApJ...863..192L, 2024ApJ...970..110G}). The eruption of the prominence is usually associated with solar flares and coronal mass ejections (CMEs; \citealt{1979SoPh...61..201M, 1987SoPh..108..383W, 2000ApJ...537..503G, 2001ApJ...559..452Z, 2002A&A...382..666H, 2003ApJ...586..562G}). If the prominence erupts completely, it is known as a full eruption. If only some parts of the prominence erupt, then it is known as a partial eruption (\citealt{2012ApJ...750...12S}). In some cases, the rise of prominence is confined by the overlying arcades. Hence, the plasma falls back towards the surface, and it is known as the failed eruption \citealt{2011RAA....11..594S, Jelnek2026}. Often, the full eruption drives the CMEs. Mostly, the eruption/disappearance episode of the prominence is preceded by an activation phase, and this activation phase (i.e., pre-eruption phase) appears in the form of various observational signatures, namely, sporadic heating events lead to fragmentary brightening and fading (e.g.,\citealt{1998SoPh..183...97O, 2008ApJ...673..611K}), deformation and helical rotation of filaments (\citealt{2011ApJ...731L...3S}), large-amplitude oscillations (\citealt{2006A&A...449L..17I, 2014ApJ...786..151S, 2014ApJ...795..130S}), increase of mass and turbulent motion at the footpoints of prominence (\citealt{2009SoPh..259...13G}), and lifting of the prominence (e.g., \citealt{2000ApJ...537..503G, 2004ApJ...602.1024S, 2009SoPh..256...57L}).\\

Since the 1970s, studies on coronal jets have attracted great interest in the field of solar physics. Various space missions in the last three decades have provided exclusive details on the initiation and evolution of the coronal jets (e.g., \citealt{Shibata1992, Yokoyama1995, Yokoyama1996, Innes1997, Shibata2007, Nisizuka2008, 2015Natur.523..437S, Jelinek2015, 2017ApJ...851...67S, 2017ApJ...849...78K, Kayshap2018, 2021MNRAS.505.5311K, Mishra2023, 2024ApJ...977..141K, 2025ApJ...994...47S}). The triggering of the solar jet is one of the important aspects of solar jet research; over the course of time, various models have been proposed for the initiation of solar jets, namely, (1) jet triggering due to the interaction of emerging flux with pre-existing coronal field (e.g., \citealt{Yokoyama1996, 2004ApJ...614.1042M, 2011ApJ...735L..43S, 2013ApJ...769..134M, 2013ApJ...769L..21A, 2014ApJ...789L..19F}), (2) jet triggering due to the kink-instability in the fan-spine magnetic topology (e.g., \citealt{2009ApJ...691...61P, 2016A&A...594A..64P, 2019ApJ...885L..11S}), and most importantly, (3) jet triggered as per the magnetic breakout model (e.g., \citealt{2016A&A...594A..64P, 2017Natur.544..452W, 2018ApJ...854..155K}). In a remarkable work, based on the morphology, \cite{2010ApJ...720..757M} has classified coronal jets into two categories, standard jets and blowout jets. Further, it is found that the eruption of the flux rope/mini filaments (or prominence) leads to the formation of the blowout jets (e.g., \citealt{2012ApJ...745..164S, 2013ApJ...769..134M, 2004ApJ...614.1042M, 2015Natur.523..437S, 2017ApJ...851...67S, 2019ApJ...885L..11S}). Now, it is almost established that the formation of the blowout jets is related to the eruption \citealt{2021RSPSA.47700217S}.\\

In some cases, the blowout jets have rotation, and this rotation of jets is due to helical field lines, which form/develop within the jet's body after the initiation of jets. (e.g., \citealt{2010ApJ...720..757M, 2011ApJ...735L..43S, 2013ApJ...769..134M, Kayshap2013b}). It is essential to understand how these helical field lines are formed. Firstly, the footpoint motions can introduce twist (i.e., adding magnetic free energy) within the fan–spine topology (i.e., a closed dome), and reconnection occurs between twisted, closed field lines and ambient, untwisted, open field lines through kink instability. Finally, the twisted field forms (e.g., \citealt{2009ApJ...691...61P}). Secondly, the reconnection between emerging twisted field lines and open field lines can also produce helical field lines (e.g. \citealt{2014ApJ...789L..19F, 2015ApJ...801...83C}). Additionally, we mention that the eruption of the prominence can show the helical motion as the twisted magnetic field of the prominence relaxes (e.g., \citealt{2009ApJ...691.1079L, 2011SSRv..158..237L, 2012ApJ...752L..22L, 2014ApJ...785L...2S, 2018ApJ...860...80P, 2019ApJ...873...22S}). \\

In this study, we have diagnosed a blowout jet-like eruption of a prominence on October 06, 2023, from 17:40~UT to 18:40~UT as seen by IRIS, AIA, and LASCO/SoHO. The prominence eruption is associated with two C-class flares and a CME. This study aims to understand the kinematics, untwisting motion, and transverse motion of the blowout jet-type prominence eruption. For this purpose, we have used data mainly from IRIS and AIA. Section 2 describes the observation and data analysis, and the observational results are explained in Section 3. The summary, discussion, and conclusion of the study are described in the last Section. 

\section{Observation and Data Analysis} \label{sec:obs}
The full-disk images of the Sun in various filters (e.g., AIA 4500~{\AA}, AIA 1600~{\AA}, AIA 1700~{\AA}, AIA 304~{\AA}, AIA 171~{\AA}, AIA 211~{\AA}, AIA 131~{\AA}, and AIA 94~{\AA}) are provided by the Atmospheric Imaging Assembly (AIA; \citealt{2012SoPh..275...17L}), which is an instrument onboard the Solar Dynamics Observatory (SDO; \cite{2012SoPh..275....3P}). These filters capture the emission from the photosphere to the corona with a spatial resolution of 0.6$"$, see  \cite{2012SoPh..275...17L} for more details about the AIA instrument. In this study, the imaging observations are used from AIA/SDO, and we have used AIA~304~{\AA}, AIA~171~{\AA}, and AIA~94~{\AA} filter observations. Next, the Interface-Region Imaging Spectrograph (IRIS) also provides high-resolution images (i.e., slit-jaw images (SJIs)) of the solar atmosphere using different filters; see \cite{2014SoPh..289.2733D} for more details. The IRIS/SJI~2796~{\AA} filter, which captures the emission from the chromosphere, is used in the present work.\\

In addition, IRIS provides the spectra from the far ultraviolet (FUV; 1331.7~{\AA} to 1407.0~{\AA}) and near ultraviolet (NUV; 2782.7~{\AA} to 2835.1~{\AA}) parts of the solar spectrum of the solar atmosphere. This FUV and NUV spectrum contains various emission and absorption spectral lines, namely, Mg~{\sc ii} k 2796.35~{\AA}, Mg~{\sc ii} h 2803.52~{\AA}, Si~{\sc iv} 1393.7~{\AA}, Si~{\sc iv} 1402.75~{\AA}, and many more. The IRIS has observed this event, and we have utilized Si~{\sc iv} 1393.75~{\AA}, Si~{\sc iv} 1402.77~{\AA}, O~{\sc iv}, 1401.16~{\AA}, and O~{\sc iv} 1399.77~{\AA} in this study. All the utilized spectral lines are optically thin lines; therefore, a single Gaussian can easily characterize these lines. We have applied a single Gaussian fit to deduce the spectroscopic parameters, i.e., peak intensity, centroid, and sigma. The total intensity (i.e., area under the curve) is obtained using the peak intensity and sigma. To convert the centroid into the Doppler velocity, we need the rest wavelength of a particular line. And, we can estimate the rest wavelength using the limb method, chromospheric method, or lamp method (\citealt{1999ApJ...522.1148P, 2024MNRAS.528.2474B}). As per the limb method, the line-of-sight (LOS) velocities are assumed to cancel each other; therefore, the line centroid at the limb can be considered the rest wavelength for that particular line (e.g., \citealt{1999ApJ...522.1148P, 2023MNRAS.526..383K, 2024ApJ...977..141K}). In this study, the limb method is used to estimate the rest wavelength of the utilized spectral lines. Further, the non-thermal velocity is calculated using iris$\_$nonthermalwidth.pro routine. This data analysis methodology is applied at each pixel and produces total intensity, Doppler velocity, and non-thermal velocity maps of the observed region, see these maps from Si~{\sc iv} line in Figure~\ref{fig:ivw_map}.\\

The Large Angle Spectrometric Coronagraph Experiment (LASCO; \citealt{1995SoPh..162..357B}) has three different coronagraphs that cover different ranges of heights above the Sun, namely, C1 (1.1 to 3 solar radii), C2 (1.5 to 6 solar radii), and C3 (3.7 to 30 solar radii). Please note that the C1 coronograph is not working at the time of our observation. Hence, we used C2 and C3 coronographs in the present study to investigate the dynamics of the associated coronal mass ejection (CMEs). All the details of the CMEs are available at this dedicated page\footnote{\url{https://cdaw.gsfc.nasa.gov/CME_list/UNIVERSAL_ver2/2023_10/univ2023_10.html}}. 
\begin{figure}
 \centering
 \mbox{
    \includegraphics[trim = 3.0cm 0.8cm 4.0cm 2.0cm, scale=1.2]{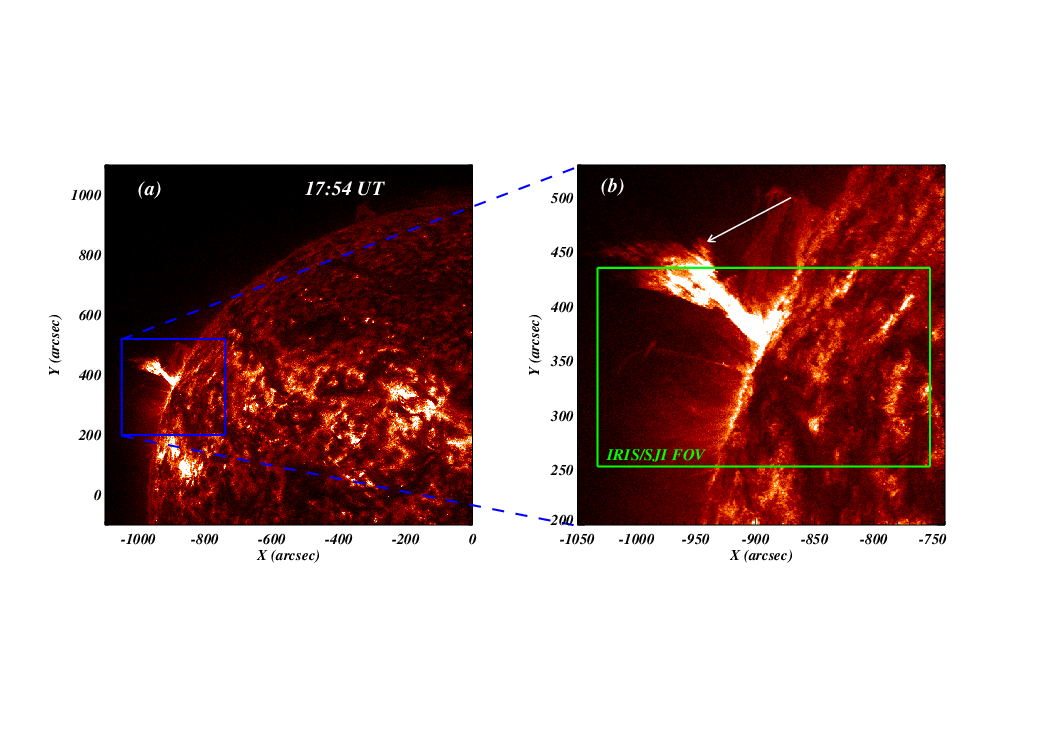}
    }
    \mbox{
    \includegraphics[trim = 1.0cm 0.0cm 1.0cm 2.0cm,scale=1.0]{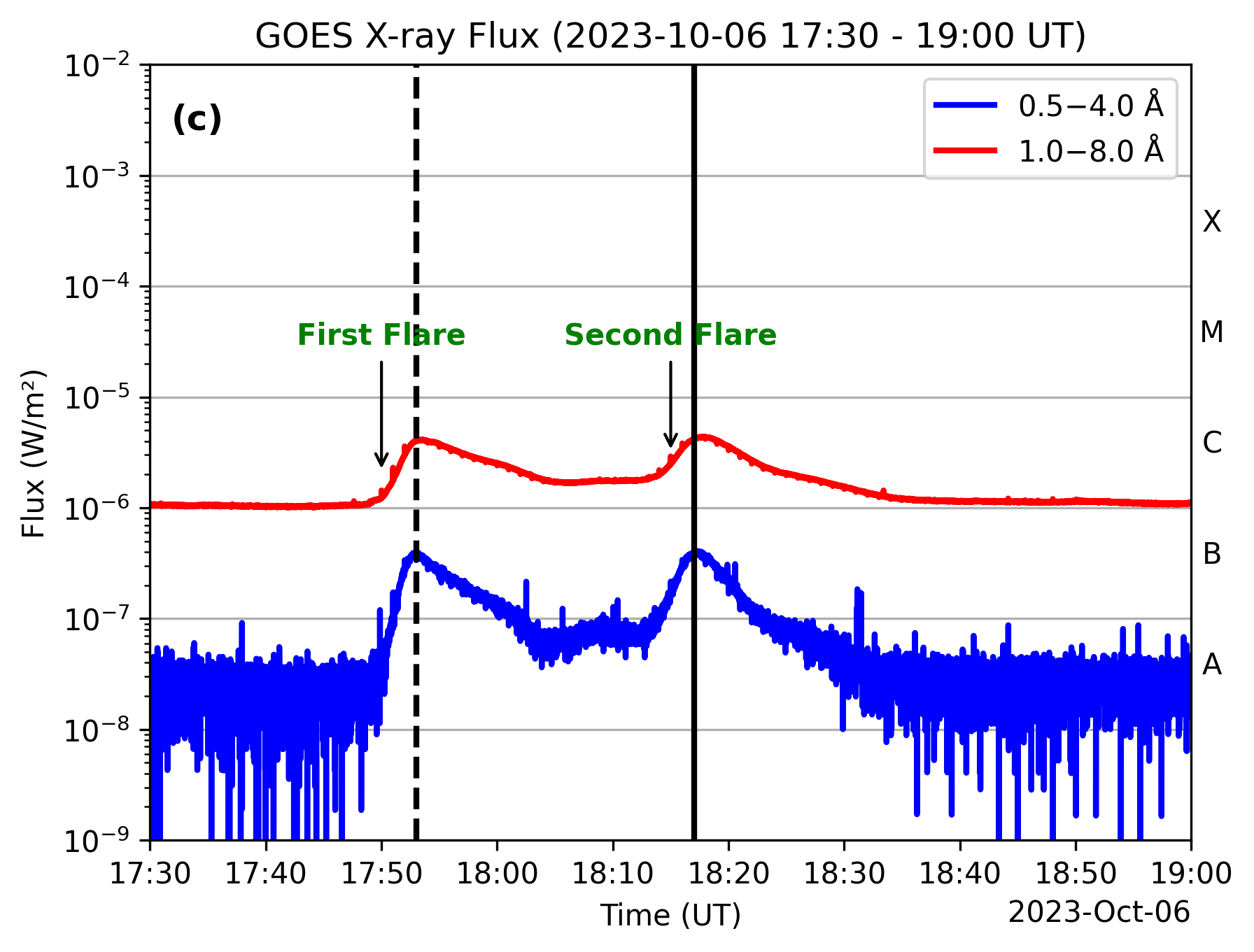}
        }
    \caption{The north-east limb of the Sun captured by AIA~304~{\AA} at 17:54~UT is shown in panel (a). The blue box represents the region where the prominence eruption occurred. This region (inside the blue box) is displayed in panel (b). The green box in panel (b) represents the field of view (FOV) of IRIS/SJI. The top part of the jet, indicated by a white arrow, was not observed by IRIS. Panel (c) displays the X-ray flux observed by GOES from 17:30~UT to 19:00~UT. The red curve corresponds to the soft X-ray (SXR) flux, and the blue curve corresponds to the hard X-ray (HXR) flux. The first peak (black dashed vertical line) and the second peak (black solid vertical line) correspond to the C4.0 and C4.2 class solar flares. The first flare initiated at 17:43~UT (marked by the black arrow) and the second flare initiated at 18:15~UT.}
  \label{fig:ref_img}
\end{figure}
Figure~\ref{fig:ref_img} is a reference image of this event. It displays a prominent eruption at the north-east limb of the Sun that occurred on October 06, 2023. The panel (a) of Figure~\ref{fig:ref_img} shows the north-east limb of the Sun, and the eruption is enclosed in a blue box. Panel (b) provides the zoomed view of the region within the blue box, and the blowout jet (or erupted prominence) is well visible. The green rectangular box in panel (b) displays the field-of-view (FOV) of IRIS. It is clearly visible that the top part of the jet (indicated by the white arrow) is not observed by IRIS, due to the limited FOV of IRIS. Panel (c) shows the soft X-ray (SXR, red curve) and hard X-ray flux (HXR, blue curve) from 17:30~UT to 19:00~UT. The two peaks occur at 17:53~UT and 18:17~UT in the SXR and HXR, and, according to the classification, the first and second peaks correspond to C4.0 and C4.2 class flares, respectively. The first flare starts at 17:43~UT and ends at 18:01~UT, while the second flare starts at 18:12~UT and ends at 18:22~UT. Please note that both time ranges lie within the time range of the current event. Moreover, as per the solar monitor\footnote{\url{https://solarmonitor.org/}}, the rise in SXR/HXR flux is due to the energetic event happening at N23E87 and N22E87, i.e., the location of our event. Hence, these things confirm that the two C-class flare occur. However, the source region is on the far side of the Sun, and we have checked Solar Orbiter and STEREO observations. But, unfortunately, the far side of the event is not captured by these observatories. 

\section{Observational Results} \label{sec:res}
\subsection{Evolution of Jet}\label{sec:jet_evol}
\begin{figure}
\centering
\includegraphics[trim = 2.0cm 1.0cm 2.0cm 0.0cm, scale=1.2]{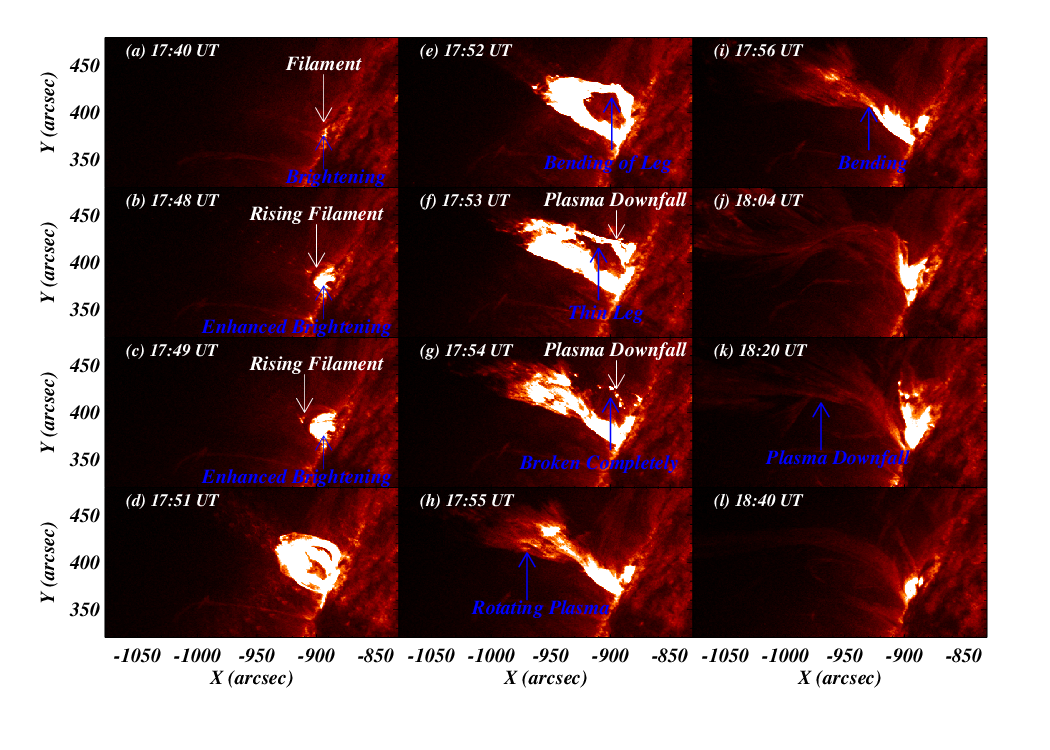}
\caption{The time evolution of the jet-like prominence eruption observed by AIA~304~{\AA} from 17:40~UT to 18:40~UT is shown here. An initial brightening and a stable filament/prominence are observed at 17:40~UT (panel (a)). The enhancement of the brightening and the rise of the filament are shown in panels (b), (c), and (d). Further, the northern leg of the prominence breaks at 17:52~UT, as indicated by the blue arrow in panel (e). As a result, the twisted jet-like structure forms, as visible in panels (g), (h), and (i). Later, the plasma falls back along the same path (panels (j) and (k)) and completely disappears at 18:40~ut (panel (l)). The animation starts from 17:39~UT to 18:41~UT. Total duration of the animation is 15s.}
\label{fig:aia304}
\end{figure}

\begin{figure}
\centering
\includegraphics[trim = 2.0cm 1.0cm 2.0cm 0.0cm, scale=1.2]{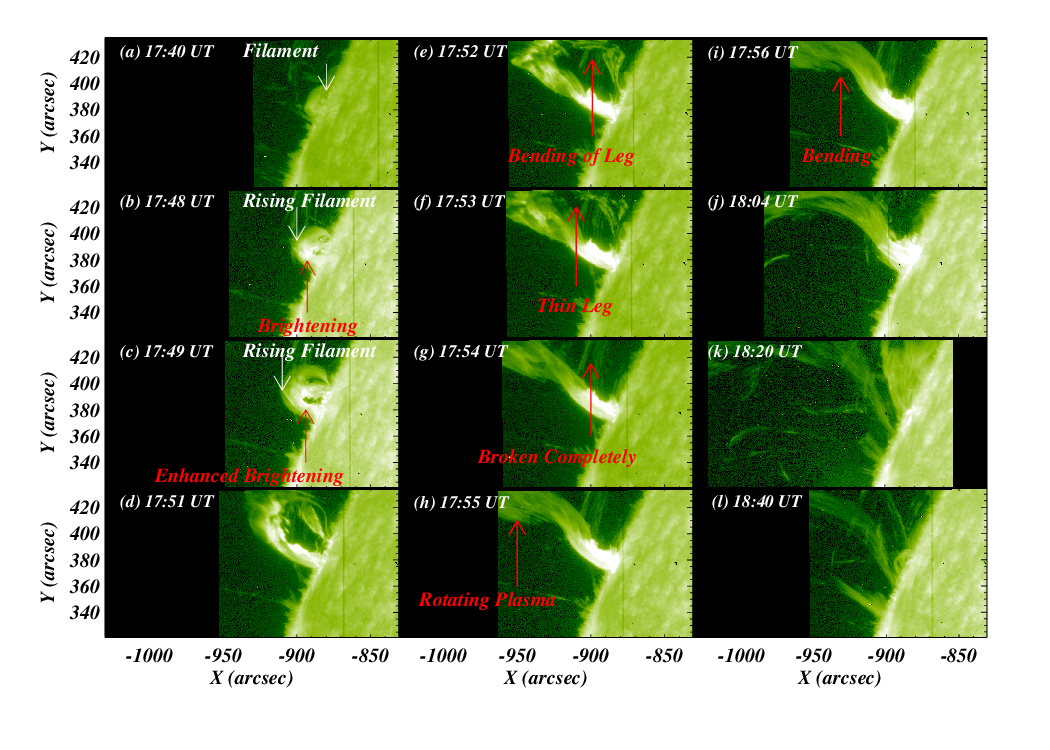}
\caption{The same signatures as Figure~\ref{fig:aia304} but observed by IRIS/SJI~2796~{\AA}.}
\label{fig:iris2796}
\end{figure}

\begin{figure}
\centering
\includegraphics[scale=1.0]{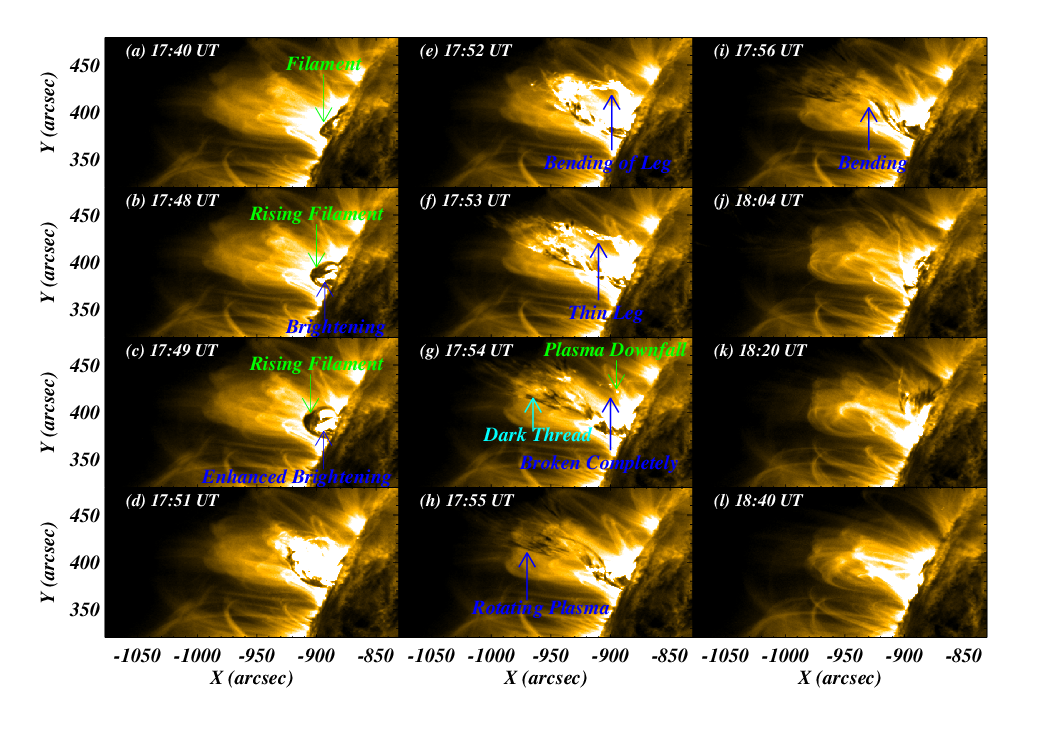}
\caption{Same as Figure~\ref{fig:aia304} but for AIA~171~{\AA} filter. Additionally, the dark thread (i.e., cool plasma threads) is visible in panels (g), as indicated by the cyan arrow. The same dark threads are visible in panels (f) and (h). The animation starts from 17:39~UT to 18:41~UT. Total duration of the animation is 15s.}
\label{fig:aia171}
\end{figure}

\begin{figure}
\centering
\includegraphics[scale=1.0]{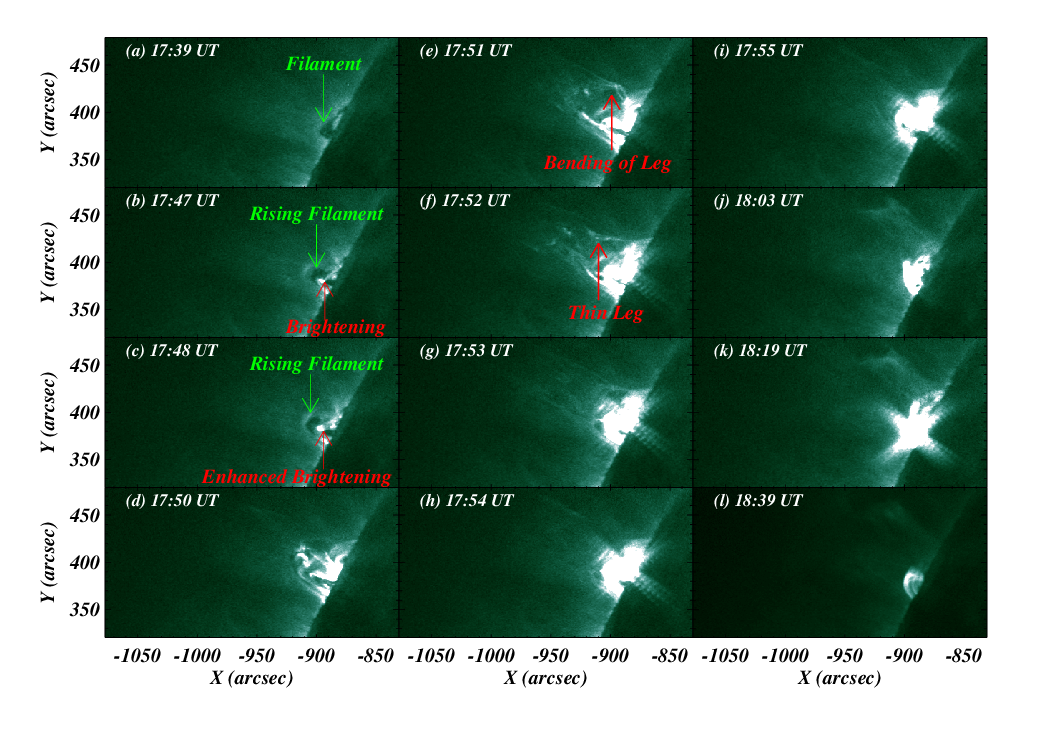}
\caption{Same as Figure~\ref{fig:aia304}, but for hot AIA~94~{\AA} filter. The animation starts from 17:39~UT to 18:41~UT. Total duration of the animation is 15s.}
\label{fig:aia94}
\end{figure}

Figure~\ref{fig:aia304} shows the evolution of the polar jet in AIA~304~{\AA} filters, which captures the emission from the transition-region (TR), i.e., plasma at a temperature of approximately 80,000 K. 
We have selected 12 key images that show the event's key observational findings, and these findings are useful for understanding the event's origin, evolution, and dynamics. In panel (a), a dark thread-like structure, indicated by the white arrow, exists around the small brightenings (indicated by the blue arrow), which is the filament/prominence. This filament existed well before the jet, i.e., around 1 hour before (not shown here). This filament/prominence is clearly visible, as a dark structure, in other high temperature filters, i.e., AIA~171~{\AA} (panel (a); Figure~\ref{fig:aia171}) and AIA~94~{\AA} filters (panel (a); Figure~\ref{fig:aia94}). While the filament/prominence appears as a bright structure in cool IRIS/SJI~2796~{\AA} (panel (a); Figure~\ref{fig:iris2796}). At T = 17:48~UT, the filament has reached higher heights (indicated by the white arrow in panel (b)). The brightenings have also increased compared to the previous time, as indicated by the blue arrow in panel (b). The filament rise and increase in brightness are also visible at T = 17:49 UT. It shows that the filament/prominence is rising over time, and its brightness is increasing over time.\\ 

At 17:51~UT, a well-developed bipole-like structure is visible (panel d), and later on, the north leg of this bipole bends as indicated by the blue arrow in panel (e). With time, the north leg is becoming thinner (panel f), and finally, it is completely broken at 17:54~UT (panel g). Some plasma falls downward after breaking the leg, which is indicated by the white arrow in panel (g) of Figure~\ref{fig:aia304}. Now, the blowout jet is clearly visible in panel (g). Just after breaking the north leg, the blowout jet plasma rotates anticlockwise, as the anticlockwise rotation is clearly visible in panels (g) and (h).  
Furthermore, we observe that the spire of the blowout jet bends at a particular location, indicated by the blue arrow in panel (i). Over time, the spire continues to bend at this location, and finally, we see the S-shaped jet spire in panel (j). Later on, the plasma falls back along the curved spire indicated by the blue arrow in panel (k).\\

In addition to the cool AIA~304~{\AA} filter, we have also investigated another cool filter from IRIS observations, i.e., IRIS/SJI 2796~{\AA}. Here, please note that the FOV of IRIS is smaller compared to AIA (see IRIS FOV in panel (b) of Figure~\ref{fig:ref_img}). In addition to the smaller FOV of IRIS, some non-observed region (i.e., black portion, which varies from frame to frame) also exists within the full FOV of IRIS. Therefore, due to these two aspects, some of the features of the event are not visible to their full extent in the IRIS observation (see the black region in each panel of Figure~\ref{fig:iris2796}). However, most of the features seen in AIA~304 are visible in this cool filter (Figure~\ref{fig:iris2796}), namely, filament/prominence (panel (a)), rise of filament (panel (b) and (c)), brightening below the filament/prominence (panels (b) and (c)), break of the north leg of the bipole (panels (e), (f), and (g)), rotating motion (panel (h)), and bending of the spire of the jet (panel (i)). Please note that filament/prominence is bright in this filter, unlike AIA~171~{\AA} (panel (a); Figure~\ref{fig:aia171}) and AIA~94~{\AA} (panel (a); Figure~\ref{fig:aia94}).\\

The Figure~\ref{fig:aia171} shows jet's evolution in AIA~171~{\AA} (i.e., relatively higher temperature filter), and it clearly shows all the important features which are mentioned in the relatively cool filters (i.e., AIA~304; Figure~\ref{fig:aia304} and IRIS/SJI 2796~{\AA}; Figure~\ref{fig:iris2796}). Here, it is important to note that the filament/prominence is distinctly visible in AIA~171~ ~{\AA} filters (see panels (a), (b), (c), and (d)). In the later stage, the dark thread also exists within the jet plasma, as indicated by the cyan arrow in panel (g). This particular observation shows the eruption of the filament associated with the formation of this blowout jet.
Lastly, the evolution of the event from the high temperature filter (i.e., AIA~94~{\AA}) is displayed in the Figure~\ref{fig:aia94}. In this filter, we see filament and its rise (panels (a), (b), and (c)), brightening at the base (panels (a), (b), and (c)), bending, and breaking of the north leg (panels (e) and (f)). But the blowout jet is almost absent in this filter, i.e., this particular blowout jet is not emitting at a higher temperature. 


\subsection{Kinematics and Fine Structure}
\begin{figure}
\includegraphics[trim = 1.0cm 0.0cm 0.0cm 1.0cm,scale=0.45]{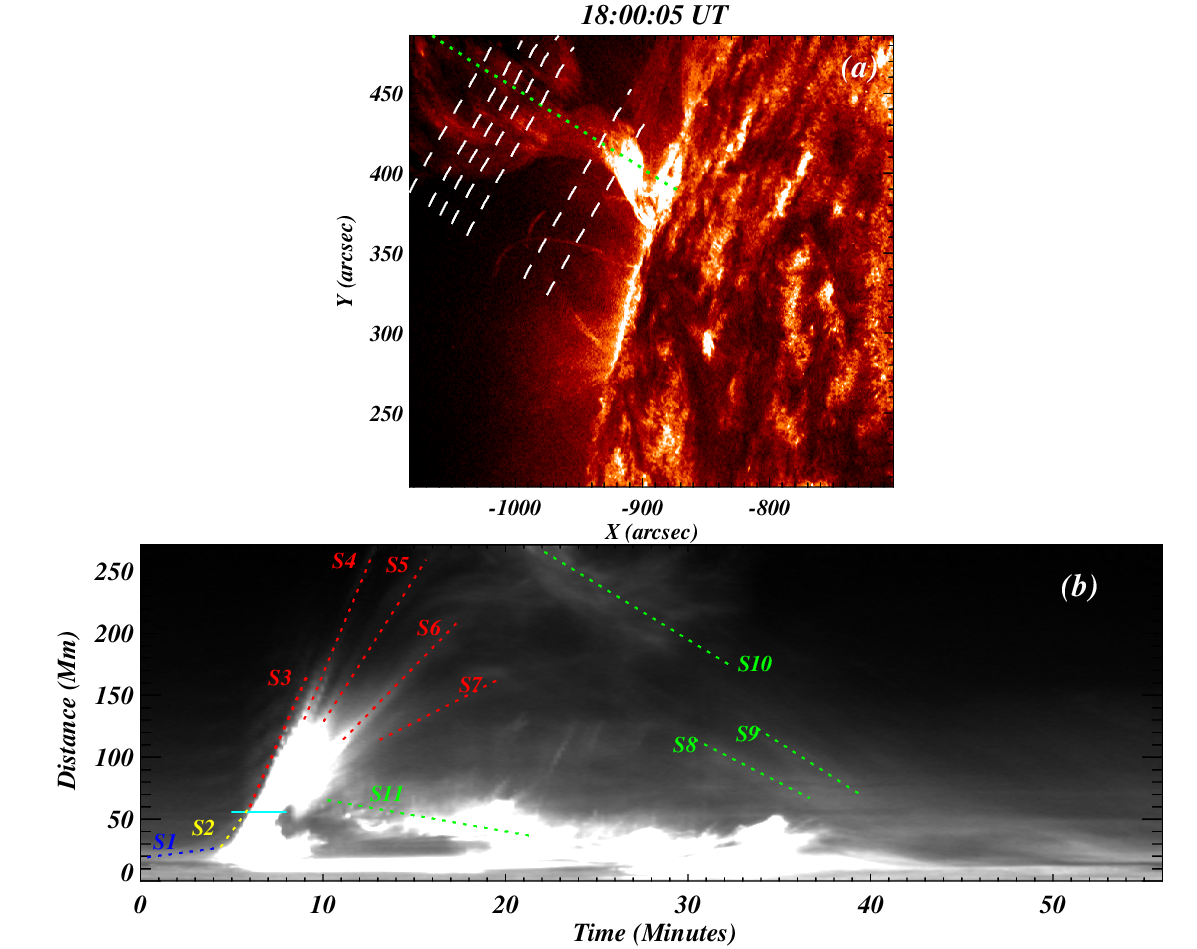}
\caption{The top panel shows the intensity image of the maximum phase of the blowout jet taken at 18.04 UT in AIA~304~{\AA} filter. One slit along the jet (vertical green dotted line) and seven slits (horizontal white dashed slits) are used to produce time-distance (TD) diagrams. The TD diagram corresponding to the slit along the jet (i.e., green dotted line) is displayed in the bottom panel. Various paths have been drawn on the TD image to estimate the speeds of various features, namely, the slow rise of prominence (blue path S1), the rapid rise phase of prominence (yellow path S2), upflow speeds of different plasma threads (red paths from S3 to S7), and downflow speeds of different plasma threads (green paths S8 to S11).}

\label{fig:td_along}
\end{figure}

The time-distance (TD) analysis is performed to estimate the kinematics of this blowout-type jet. Panel (a) of Figure~\ref{fig:td_along} shows the AIA~304~{\AA} intensity map from the maximum phase of the jet (i.e., time = 18:00~UT), and further, one slit along the blowout jet (i.e., green dashed slit) is drawn. Next, seven slits are drawn across the jet (see white dashed slits), which cover its full vertical extent. Two slits are located below the bended spire, i.e., these two slits are close to the base. While the other five slits are placed in the wide spire region of the blowout jet, i.e., above the bended spire.\\

The panel (b) of Figure~\ref{fig:td_along} shows a TD image deduced using a green slit along the jet. To produce this TD image, we have considered 80 pixels in the transverse directions, i.e., 40 pixels above and below at each location of the green slit. This horizontal extent at each point covers almost the full width of the jet. This TD image is loaded with various important information, and we explain it in chronological order.
\begin{table}[hbt!]
\centering
\caption{Various features of the event and their velocity}
\label{tab:jet_velocity}
\begin{tabular}{ |c|c|c|c| }
 \hline
 Sr. No &  Feature  & Slit/Path ID & Velocity (km/s) \\ 
 \hline
 1. & Filament Slow Rise  & S1 & 32.54 \\ 
 \hline
 2. & Filament Fast Rise  & S2 & 337.91 \\ 
 \hline
 3. & Jet Upflow (path 1) & S3 & 563.19 \\
 \hline
 4. & Jet Upflow (path 2) & S4 & 593.46 \\
 \hline
 5. & Jet Upflow (path 3) & S5 & 387.97 \\
 \hline
 6. & Jet Upflow (path 4) & S6 & 252.23 \\
 \hline
 7. & Jet Upflow (path 5) & S7 & 124.92 \\
 \hline
 8. & Jet Downflow (path 1) & S8 & 125.63 \\
 \hline
 9. & Jet Downflow (path 2) & S9 & 157.83 \\
 \hline
 10. & Jet Downflow (path 3) & S10 & 149.98 \\
 \hline
 11. & Jet Downflow (path 2) & S11 & 43.04 \\
 \hline
\end{tabular}
\end{table}

\begin{itemize}
\item Firstly, we have seen the slow rise phase of prominence, and the blue slit is drawn along the slow rise phase of prominence (see S1 blue path). The slow rise velocity of the prominence is around 32 km/s.
\item  The fast rise phase of prominence occurs later on, which is outlined by the yellow dashed path (see S2 yellow path). The fast rise velocity of the prominence is around 338 km/s. 
\item After around 12 minutes from 17:40~UT, one leg of the prominence starts breaking, and this particular location is shown by the horizontal cyan solid line. And, this is the time when the blowout jet starts to form. 
\item The blowout jet is composed of many different plasma threads, and they are moving at different speeds. We have drawn five different paths (see S3 to S7 red paths), and estimated the upflow speeds of plasma. The upflow speeds for path S3, S4, S5, S6, and S7 are 563, 593, 387, 252, and 125 km/s, respectively.   
\item Similar to the upflow speed, we estimated the downflow speed of plasma from different threads, see S8 to S10 green paths. The downflow speeds for S8, S9, and S10 are 125, 157, and 149 km/s. 
\item The last green path (i.e., S11) is a path drawn along the plasma, which falls towards the surface after the breaking of the north leg of the prominence. The plasma downfall speed is around 43 km/s.
\end{itemize}

All the values of velocities (i.e., slow and fast rise of prominence, upflow speeds, and downflow speeds) are mentioned in the table~\ref{tab:jet_velocity}.
\begin{figure}
\centering
\includegraphics[scale=0.4]{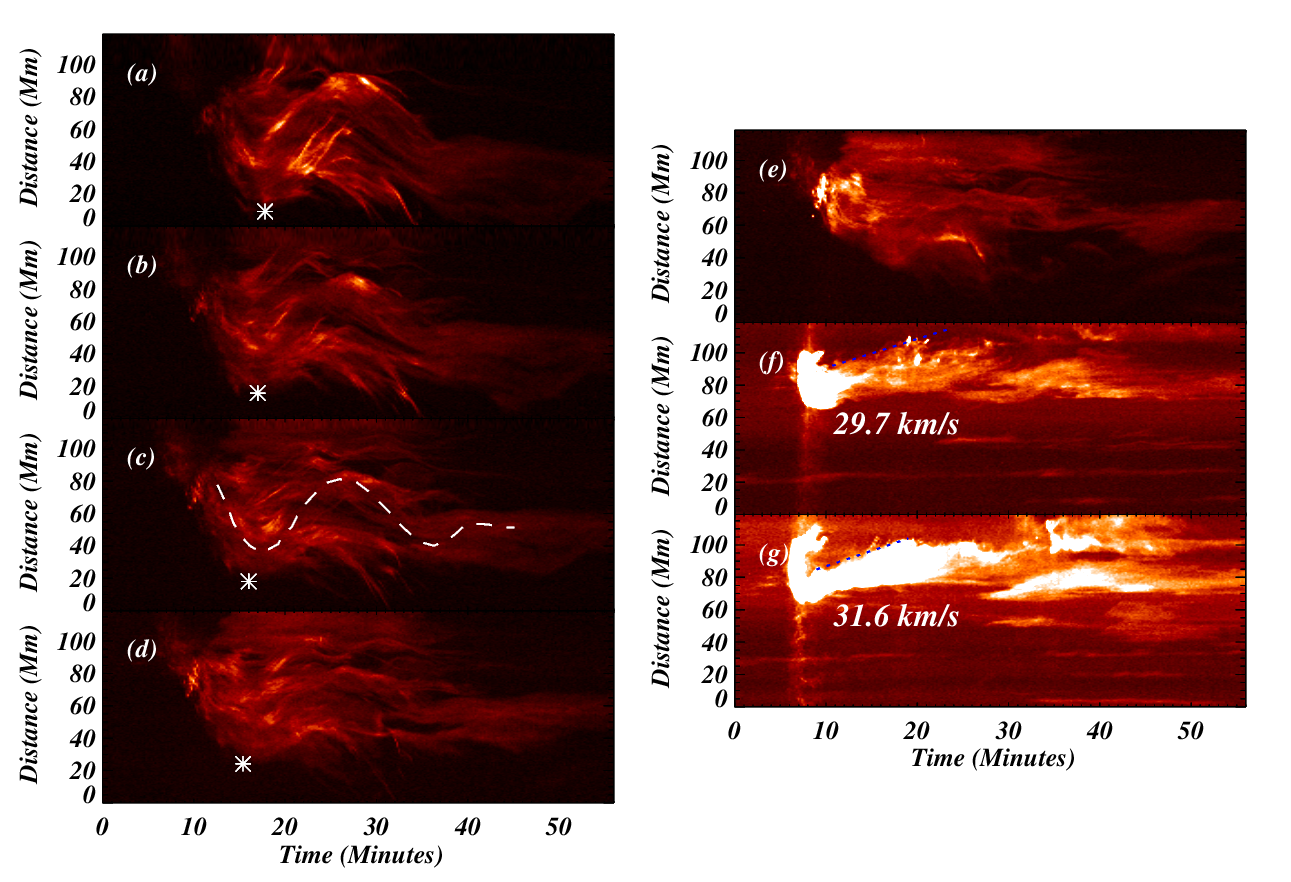}
\caption{Figure shows various TD images corresponding to the slits across the jet (see white dashed lines in Figure~\ref{fig:td_along}(a). The TD image in panel (a) corresponds to the top white slit in Figure~\ref{fig:td_along}(a). While the TD image in panel (g) corresponds to the bottom-most white slit in Figure~\ref{fig:td_along}(a). The TD image from the last two slits (i.e., white slits below the bended location) shows that the spire of the blowout jet expands at a speed of around 30 km/s. The TD image from the top 5 slits (i.e., above the bended spire) shows transverse oscillations, and the main oscillation is manually shown by a white dashed path in panel (c). The trough position (indicated by the white asterisk in panels from (d) to (a)) is used to estimate the phase speed.}   
\label{fig:td_across}
\end{figure}
Further, we have deduced TD images from the slits across the jets (see white dashed slits, Figure~\ref{fig:td_along}), and they are displayed in Figure~\ref{fig:td_across}. The TD images shown in panels (a), (b), (c), (d), (e), (f), (g), and (h) correspond to the top white slit to the bottom white slit, respectively. Here, it should be noted that Jet's spire is well collimated near the base, i.e., below the prominent bend. The bottom two slits are located in the well-collimated spire of the blowout jet. The TD maps from the bottom two slits (i.e., panels (g) and (h)) show the spire of the jet expands with time, see the blue dashed paths. The expansion speeds are 32 and 30 km/s for the bottom slits; the width of the spire expands from 15 Mm to around 30 Mm. 

\subsection{Transverse Oscillations}
The jet's spire appears as a wide funnel above the prominent bend. The top five slits are located in the wide spire of the jet (panel (a); Figure~\ref{fig:td_along}). The jet's spire, above the bend, also has significant untwisting motion. Therefore, the TD images from the top four slits clearly show an oscillation-like pattern (see panels (a) {--} (d) of Figure~\ref{fig:td_across}). Manually, this wave pattern is shown by a white dashed line in panel (c) of Figure~\ref{fig:td_across}. Visually, it is also clear that oscillations shift from left to right from lower height (panel (d) of Figure~\ref{fig:td_across}) to higher height (panel (a) of Figure~\ref{fig:td_across}), i.e., oscillations occur first at lower height and later at higher heights. In other words, we can say that oscillation is propagating upward, and we estimate the phase speed of this oscillation. To know that phase speed, we selected one particular location (i.e., trough point of oscillation), and it is marked by the asterisk symbol in panels (a) to (d) of figure~\ref{fig:td_across}. Now, we know the height difference between the top white slit and the fourth white slit, shown in panel (a) of Figure~\ref{fig:td_along}, and also we know the corresponding time of this location at the first and fourth white slit. Using this information, we found that the phase speed is around 267 km/s.\\

We know that blowout-type jets have various plasma threads within their bodies (\citealt{2010ApJ...720..757M, 2013ApJ...769..134M}). As already mentioned, the jet plasma shows oscillation as a whole (see white dashed path in panel (c); Figure~\ref{fig:td_across}). Apart from this oscillation, we have noticed individual oscillations in various plasma threads.
\begin{figure}
\centering
\includegraphics[scale=0.4]{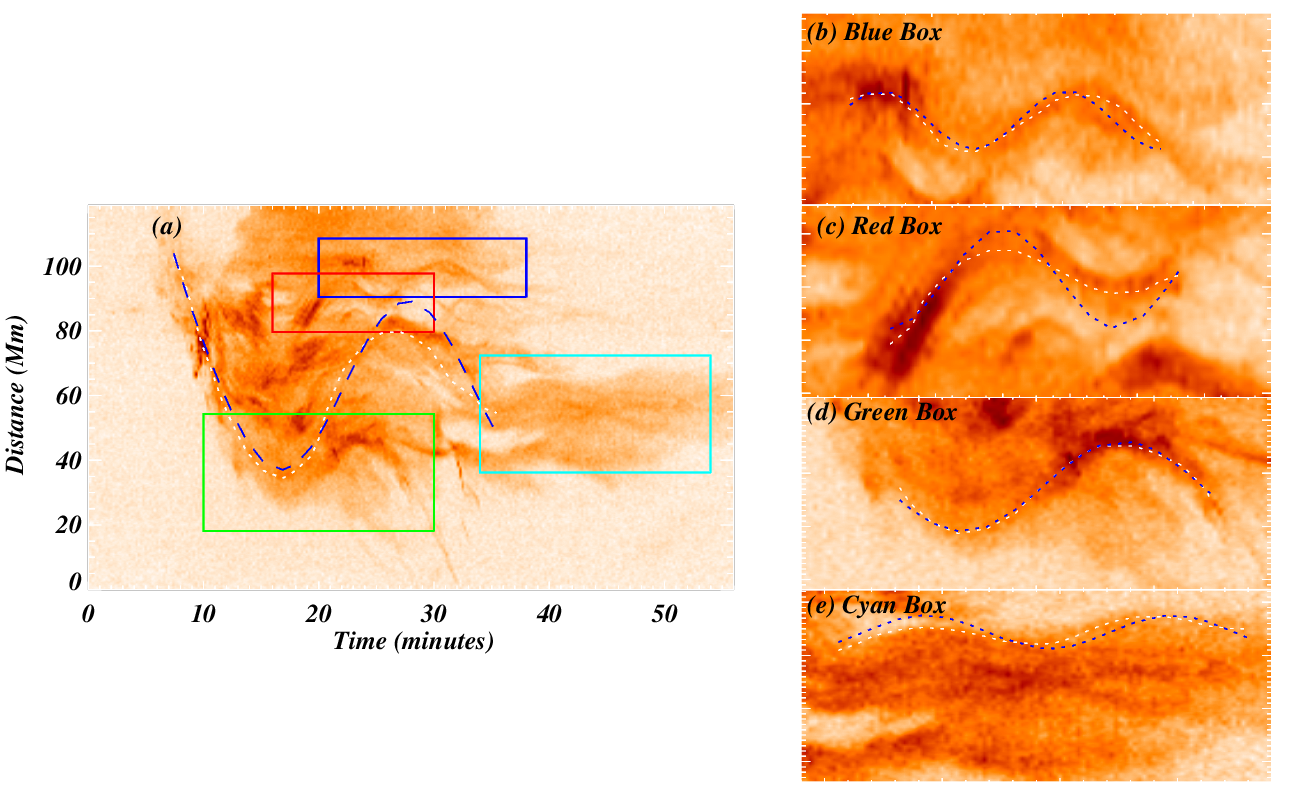}
\caption{The TD image in panel (a) shows the transverse oscillation, and this TD image corresponds to panel (c) of Figure~\ref{fig:td_across}. The main oscillation is tracked manually (white dashed path), and further fitted using the equation~\ref{eq2} (blue dashed curve). Further, the individual oscillations are spotted in four different plasma threads, and the region of these oscillation are outlined by the blue, red, green, and cyan color boxes. And, the zoomed view of these boxes is shown in panels (b), (c), (d), and (e), respectively. Further, the transverse oscillations are tracked manually, see the white dashed paths in panels (b) to (e). Next, the manually tracked paths are fitted using the equation~\ref{eq1}, and the fitted curve is shown by blue dotted lines in panels (b) to (e).} 
\label{fig:osci}
\end{figure}
The TD image (in reverse color) displayed in panel (a) of Figure~\ref{fig:osci} is the same as shown in panel (d) of Figure~\ref{fig:td_across}, i.e., the TD image from the fourth slit. We have distinctly identified the oscillations in four different plasma threads, which are enclosed by rectangular boxes, see green, red, blue, and cyan rectangular boxes in (a) of Figure~\ref{fig:osci}. The zoomed view of the region enclosed by the blue box is displayed in panel (b) of Figure~\ref{fig:osci}, and we clearly see the oscillation in this particular plasma thread. Firstly, we manually drew the path on the oscillation, indicated by the white dashed line. To fit this oscillation, we have used the simple wave equation.
\begin{equation}
S_{}(t) = A_{1} \sin(\omega t) 
\label{eq1}
\end{equation}
Here, $S(t)$, $A_{1}$, $\omega$, and $t$ are the distance perpendicular to the jet, amplitude, angular frequency, and time, respectively. The fitted curve is shown by a blue dashed line in panel (b). Similarly, the zoomed views of red, green, and cyan rectangular boxes are shown in panels (c), (d), and (e) of Figure~\ref{fig:osci}. The oscillations are distinctly visible in all the panels. And, the white dashed line is a manually drawn path, while the blue dashed line is the fitted curve (see equation~\ref{eq1}). With the help of the fitted wave equation, we estimated the amplitude and period of the oscillations, and they are tabulated in Table~\ref {tab:osci_para}. The amplitude varies from 2.74 Mm to 8.52 Mm, while the period varies from 406 s to 811 s. Further, we estimate the transverse velocity near the equilibrium (mean) position, and the values are tabulated in Table~\ref {tab:osci_para}. The transverse velocities at the mean position are 37.71$\pm$1.15 (blue box),  66.95$\pm$5.57 (red box), 64.92$\pm$1.44 (green box), and 30.10$\pm$0.96 (cyan box). Lastly, the lifetime ranges of these oscillations are added in the last column of this Table~\ref{tab:osci_para}.
\begin{table}[hbt!]
\centering
\caption{Kinematics of the oscillations}
\label{tab:osci_para}
\begin{tabular}{ |c|c|c|c|c|c|c| }
 \hline
 Sr. No & Oscillation ID & Amplitude (Mm)& Time Period (s) & Transverse Velocity (km/s) & Time-Range (Minutes) \\ 
 \hline
 1. & Blue Box & 2.74 & 425.54 & 37.71$\pm$1.15 & 20{--}40\\
 \hline
 2. & Red Box  & 4.72 & 406.11 & 66.95$\pm$5.57 & 15{--}30\\
 \hline
 3. & Green Box & 8.52 & 811.48 & 64.92$\pm$1.44 & 10{--}30\\
 \hline
 4. & Cyan Box  & 3.13  & 628.75 & 30.10$\pm$0.96 & 33{--}55 \\
 \hline
 5. & Main Oscill. & 26.19  & 1332.88 & 126.18$\pm$7.27 & 07{--}35 \\
 \hline
\end{tabular}
\end{table}
In addition to the oscillations in the individual plasma threads, we have tracked the main oscillation in panel (a) of Figure~\ref{fig:osci}, see the white dashed (manual path). The amplitude of the oscillations in the individual plasma threads is not changing much with time (see oscillations in panels (b), (c), (d), and (e) of Figure~\ref{tab:osci_para}), therefore, we have treated them as non-decaying oscillations. However, the main oscillation is weakening with time, i.e., a damped oscillation. Hence, the main oscillation is characterized by the combination of exponential and sine functions, as displayed in the equation~\ref{eq2}.   
 \begin{equation}
 S_{}(t) = A_{0} \mathrm{e^{-bt}} A_{1} \sin(\omega t) 
 \label{eq2}
 \end{equation}
Here, $A_{0}$ and $b$ are the constant and exponent related to the exponential function. While $A_{1}$, $\omega$, and $t$ are the same parameters as used in Equation~\ref{eq1}. Using this equation, the main oscillation is fitted, see the blue dashed path in panel (a) of Figure~\ref{fig:td_across}. We have estimated the amplitude, period, and transverse velocity, and time range for this main oscillation, and all these parameters are tabulated in the last row of Table~\ref{tab:osci_para}. The amplitude, period, and transverse velocity of this main oscillation are 26.19 Mm, 1332.88 s, and 126.18$\pm$7.27 km/s, respectively.\\

Next, we would like to understand the physical relationship between the four individual oscillations in plasma threads and the main oscillations. The oscillations in the green, red, blue, and cyan boxes start at 10, 15, 20, and 33 minutes after the main oscillation, and their amplitudes are 8.52, 4.72, 2.74, and 3.13 Mm, respectively. Hence, the amplitude decreases as per the starting time of the oscillation. Next, the transverse velocities of oscillations in green, red, blue, and cyan boxes are 65, 66, 38, and 30 km/s, respectively. It means, similar to the amplitude, the transverse velocities also decrease as per the starting time of the oscillation. The main oscillation is decaying with time; therefore, the energy of this main oscillation is decaying. The individual oscillations in plasma threads that occur later have less amplitude and transverse velocity than the oscillations that occur earlier. It means the amplitude and transverse velocity of individual oscillations are also decreasing as the main oscillation. It shows that the source of these individual oscillations is consistent with the main oscillations. The oscillation periods for green, red, blue, and cyan boxes are 811, 406, 425, and 629 seconds; this means that the oscillation period increases, as per the starting time of the oscillations (except the green box oscillation).

\subsection{Spectroscopic Analysis} \label{sect:spectra}
\begin{figure}
\centering
\includegraphics[trim = 1.0cm 4.0cm 0.0cm 1.5cm, scale=1.1]{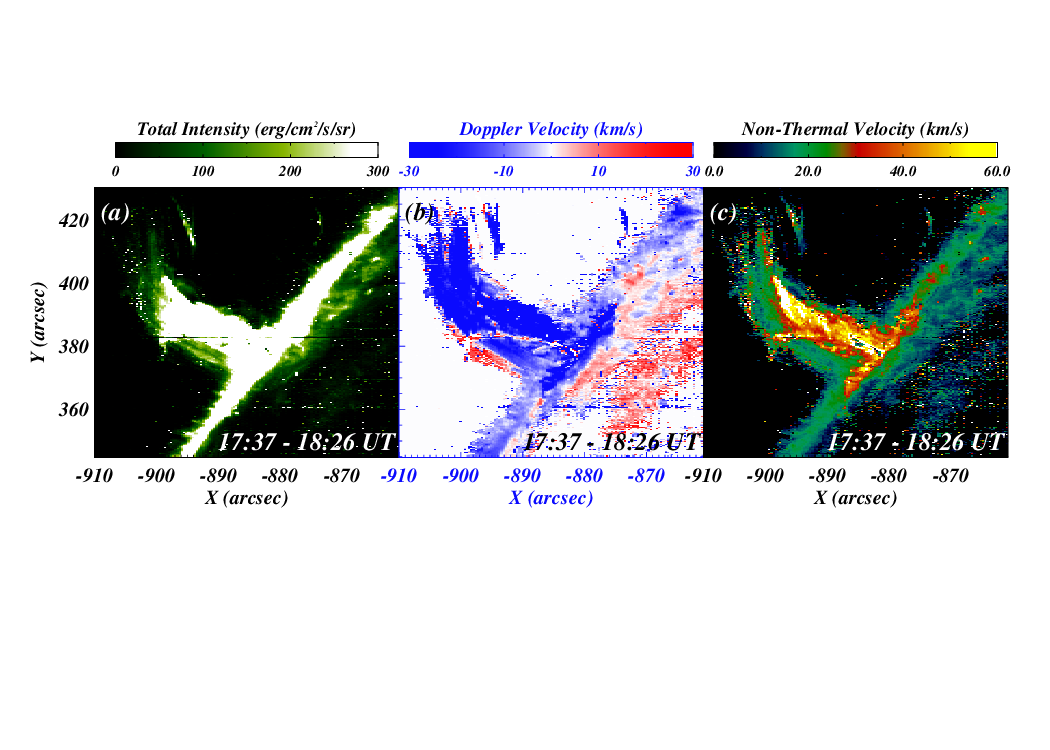}
\caption{Panel (a), (b), and (c) show the spectral intensity, Doppler velocity, and non-thermal velocity maps of the observed region, respectively. These maps are obtained by fitting a Gaussian function to the Si~{\sc iv} profiles for each pixel of the observed region.}
\label{fig:ivw_map}
\end{figure}

\begin{figure}
\centering
\includegraphics[scale=0.40]{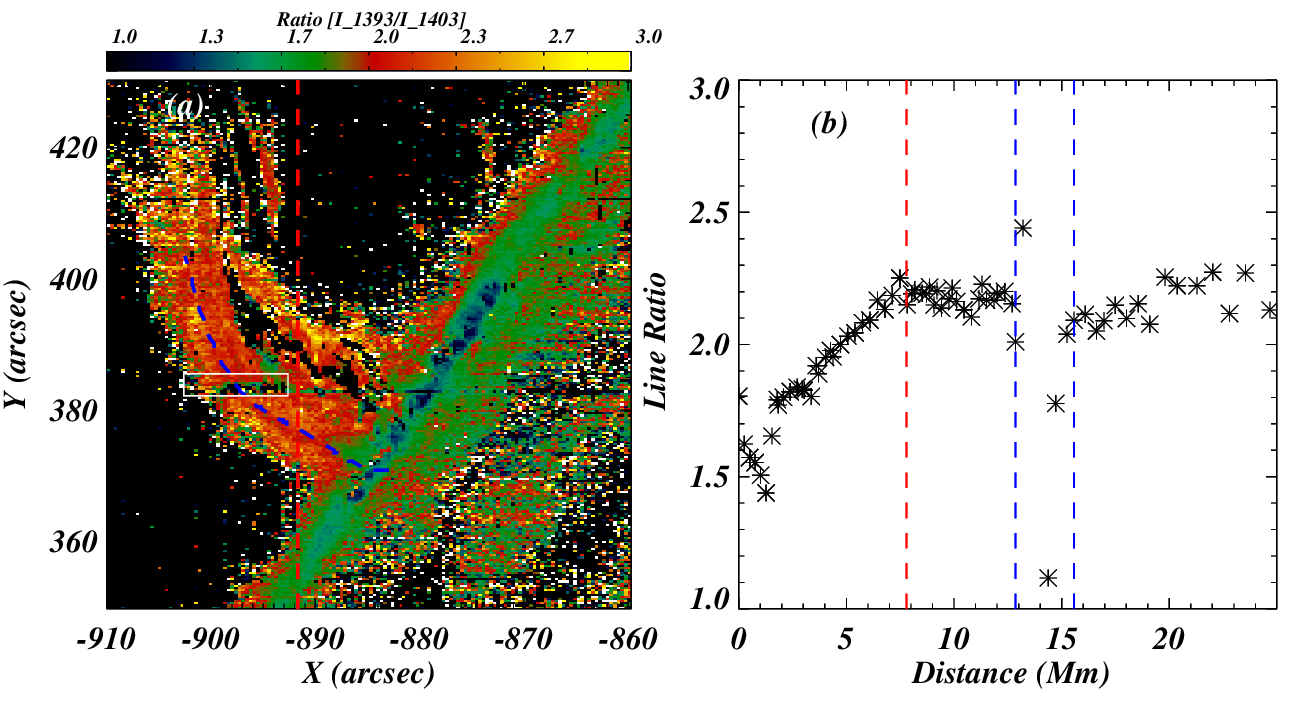}
\includegraphics[scale=0.40]{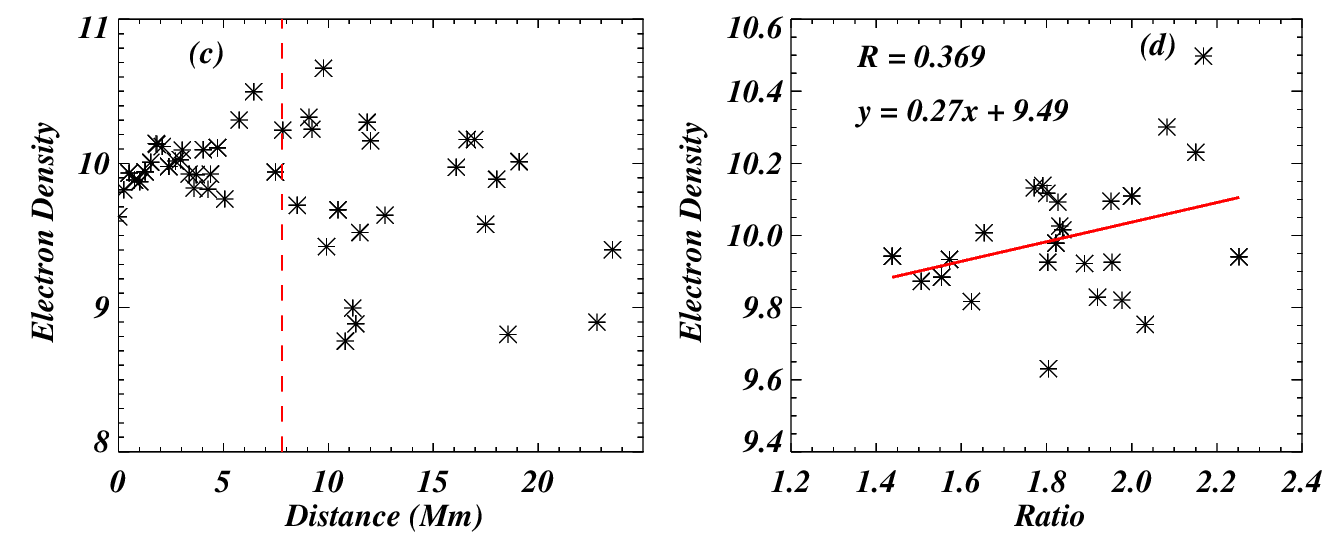}
\caption{Panel (a) shows the intensity ratio of Si~{\sc iv} (i.e., 1403/1399) map of the observed region. Most of the locations within the jet region have the ratio value close to the theoretical ratio value (i.e., 2). The path is drawn along the jet (blue dashed path in panel (a)), and the ratio along this path is shown in panel (b). The ratio value starts around 1.5 (at the limb) and increases further, reaching close to 2 at a height of about 7.8 Mm. This particular location is indicated by red vertical dashed lines in panels (a) and (b). After this height, the ratio becomes almost constant at around 2. The region enclosed by blue vertical lines in panel (b) belongs to the region where the spectra were of the poor quality (i.e., not usable spectra), and the same region is outlined by a white rectangular box in panel (a). Further, panel (c) shows the electron density along the jet (i.e., along the blue slits), and the correlation of line ratio with electron density, below the height of 7.8 Mm (i.e., red vertical slit), is displayed in panel (d).}
\label{fig:ratio_density}
\end{figure}

The spectroscopic analysis is an important tool to understand the physical conditions (e.g., plasma flows, non-thermal motions, opacity, and electron density) in any feature/activity of the solar atmosphere (e.g., \citealt{2015SoPh..290.2889K, 2017SoPh..292..108K, 2018LRSP...15....5D,2018ApJ...864...21K, 2019ApJ...874...56R, 2021ApJ...906..121K, 2023MNRAS.526..383K, 2024MNRAS.528.2474B, 2024MNRAS.529.3424H, 2024Ap&SS.369...61B, 2024ApJ...977..141K, 2025Ap&SS.370...13B}). Using the Si~{\sc iv} 1393.75~{\AA} line provided by IRIS, the total intensity, Doppler velocity, and non-thermal velocity maps of the observed region are displayed in the panels (a), (b), and (c) of Figure~\ref{fig:ivw_map}, respectively. The jet structure is distinctly visible in all three panels of Figure~\ref{fig:ivw_map}. The majority of the jet plasma is blue-shifted in panel (b); it means the plasma is up-flowing. Although a small fraction near one edge is showing redshifts. The existence of opposite Doppler shifts at the edges of solar jets is a typical signature of rotating motion of jet plasma (e.g.,\citealt{2015ApJ...801...83C, Kayshap2018, 2021MNRAS.505.5311K}). Hence, the spectral profiles across the jet (i.e., from the lower edge to the upper edge) shift from red (blue)-shifted to blue (red)-shifted profiles (e.g., \citealt{2015ApJ...801...83C, 2018ApJ...854..155K}), or at least the Doppler velocity decrease (increase) from one edge to other (e.g., \citealt{2014PASJ...66S..12Y, Kayshap2018}). We have investigated the variation in Doppler shifts across the jet at various X-positions, but we could not find a Doppler velocity pattern consistent with rotational motion. Additionally, the opposite Doppler shifts at the edges should cover a significant spatial extent. However, the present observation shows the red-shifted region covers a very small portion of the jet's area, and that too fades quickly in time. Hence, we confirm that a small redshift patch on the edge of the jet does not represent the rotational motion of the plasma. Rather, this redshift patch indicates that some plasma is falling towards the solar surface. The non-thermal velocity map exhibits extremely high values in the jet region compared to the surrounding region, indicating that the Si~{\sc iv} spectral line is broad in the jet region.\\ 
Next, we have estimated the Si~{\sc iv} line intensity ratio (i.e., I$_{1394}$/I$_{1403}$) for the observed region. This line ratio is important to know whether Si~{\sc iv} lines are forming in optically thick or thin conditions (e.g., \citealt{2020ApJ...894..128T, 2021MNRAS.505.5311K, 2024MNRAS.528.2474B, 2024Ap&SS.369...61B, 2025Ap&SS.370...13B}). The theoretical line ratio of these Si~{\sc iv} is 2, and if the observed ratio is 2, then the lines are forming in optically thin conditions, and if the ratio is less than 2, then the line is forming in optically thick conditions. The ratio map is shown in panel (a) of Figure~\ref{fig:ratio_density}. The line ratio is lower than 2 near the limb and slightly below, i.e., the Si~{\sc iv} line is forming in optically thick conditions. The line ratio increases as we move away from the limb towards the main body of the polar jet. 
To show it very clearly, we have chosen a path along the jet, see the blue dashed path in panel (a) of Figure~\ref{fig:ratio_density}. Further, the averaged ratio (i.e., averaged over 4 pixels in the vertical direction at each position) along the length of the jet (i.e., blue dashed path) is displayed in panel (b). The blue-dashed path starts below the limb. Firstly, the line ratio increases until the distance reaches approximately 7.8 Mm, and this particular location is marked by the vertical red-dashed lines in panels (a) and (b). After this, the line ratio is around 2 and does not change with height. Please note that the ratio shows very different values between the vertical blue dashed lines in panel (b). Here, please note that the area outlined by a white rectangular box in panel (a) belongs to the region of poor quality spectra, i.e., the spectra were not usable here. Therefore, this leads to an incorrect estimation of the line ratio. Hence, we have outlined line ratios belong to poor quality spectra by the blue dashed lines in panel (b).\\

Next, we estimated the electron density using the O~{\sc iv} density sensitive lines, i.e., O~{\sc iv}~1401.16~{\AA} and O~{\sc iv} 1404.81~{\AA} (e.g., \citealt{2016A&A...594A..64P}). The estimated electron density along the chosen path (i.e., blue dashed path shown in panel (a) of Figure~\ref{fig:ratio_density}) is displayed in panel (c) of Figure~\ref{fig:ratio_density}. Please note that we have removed the density points from the bad region, i.e., the region outlined by the white rectangular box in panel (a) of Figure~\ref{fig:ratio_density}. The electron density increases from the limb up to a height of 7.8 Mm, i.e, till the height indicated by the red dashed line in panel (c). Please note that, similar to electron density, line ratio also increases till the same height (i.e., 7.8 Mm) from the limb (panel (b)). But later, the electron density decreases, which is quite obvious. Lastly, we have checked the correlation between line ratio and electron density up to the distance of 7.8 Mm, i.e., for the correlation, the line ratio and electron density values are only taken from the left of the red-dashed vertical lines in panels (b) and (c) of Figure~\ref{fig:ratio_density}. The correlation is displayed in panel (d) of Figure~\ref{fig:ratio_density}, and the line ratio is positively correlated with the electron density. The Pearson's coefficient is around 37\% for this correlation. 

\subsection{Coronal Mass Ejection}
\begin{figure}
\centering
\includegraphics[trim = 2.0cm 2.0cm 0.0cm 1.0cm,scale=1.30]{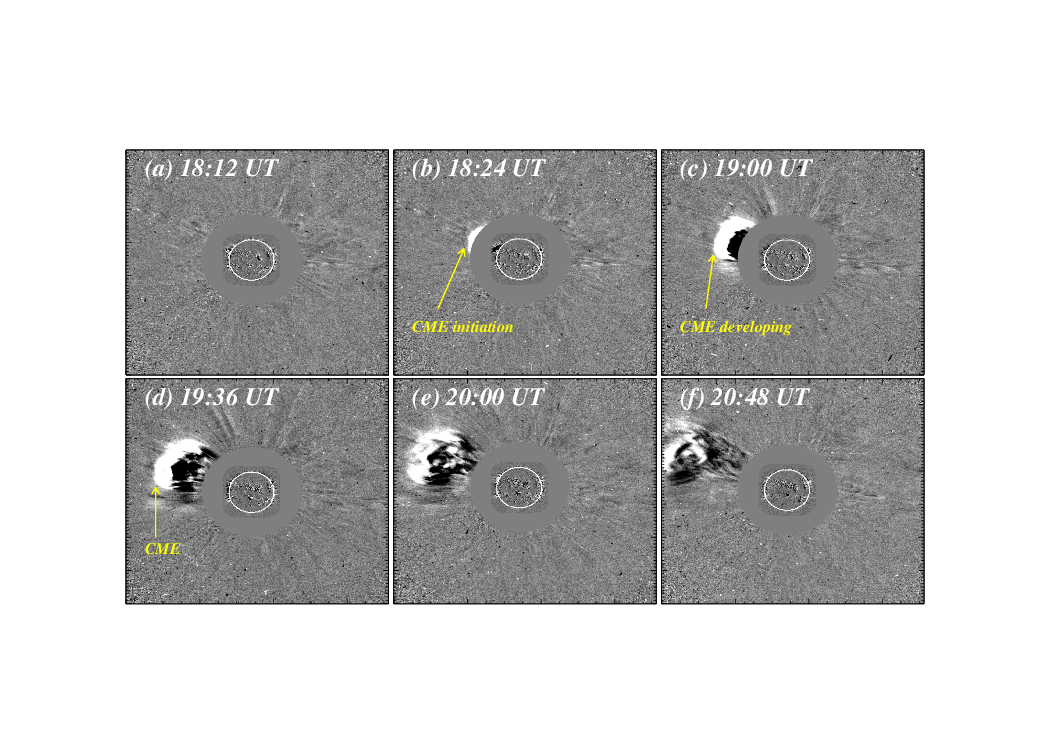}
\caption{Figure shows the evolution of CME, as observed by C2 coronograph of LASCO. The CME appears around 18:24~UT, as indicated by the yellow arrow in panel (b). Further, the CME evolves with time as displayed from panels (d) to (f).}
\label{fig:cme_evol}
\end{figure}

The LASCO/SoHo observations reveal that the event is associated with the coronal mass ejection (CME). The CME is visible in the C2 and C3 coronographs, and Figure~\ref{fig:cme_evol} shows the evolution of the CME from the C2 coronograph. The very first signature of CME appears around 18:24~UT (panel (b); Figure~\ref{fig:cme_evol}).
Here, it should be noted that prominence eruption starts around 17:52~UT, i.e., bending and breaking of one leg of prominence (Section~\ref{sec:jet_evol}). We know that the C2 coronograph provides images from 1.5 solar radii; therefore, the CME appears after around $\sim$~32 minutes (i.e., 18:24~UT). And, in the consecutive images, the CME is continuously developing, indicated by yellow arrows in panels (c) and (d). Finally, we see a semicircular CME at 19:36~UT (panel (d)). Please note the dark region behind the bright front of the CME is the coronal dimming. The CME eruption opens the magnetic field, which allows the plasma to escape easily. As a result, it creates a region of lower density and lower temperature behind the CME, which appears as a dark region, i.e., coronal dimming. In the next panels, it is well visible that the CME is moving away from the Sun with time (panels (e) and (f)). Finally, the CME disappears completely around 23:40~UT from the FOV of the C2 coronograph. However, the CME was visible in the C3 coronograph.\\
\begin{figure}
\centering
\includegraphics[scale=0.32]{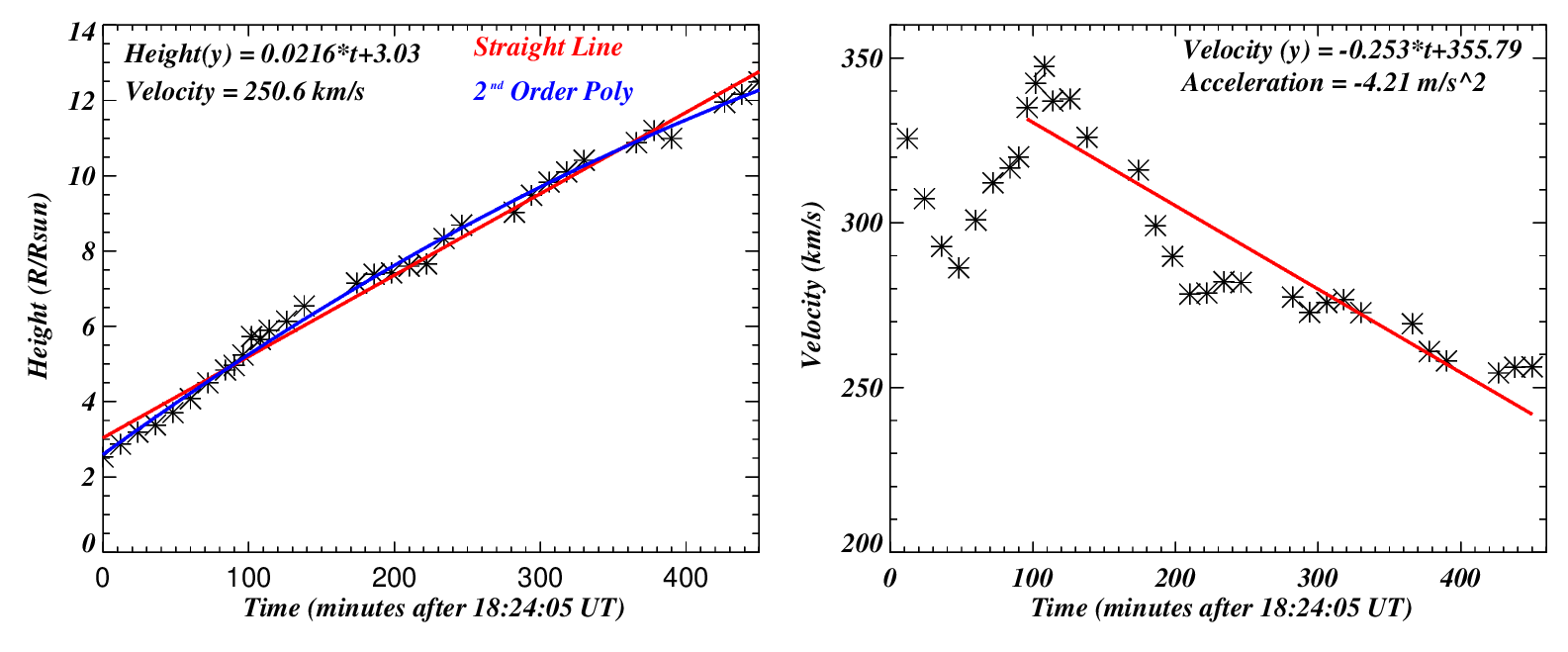}
\caption{Left panel: the height of CME against the time. The red and blue curves are the line and 2$^{nd}$ order polynomial fit on the data points. The linear fit equation is mentioned at the top of the panel, and using the slope, we found that velocity is around 250 km/s. The right panel shows the velocity of the CME with time. Firstly, the velocity increases, and later it decreases. The deceleration phase is fitted with a straight line (see red line). Using this, the deceleration is estimated, which is $-$4.21 km/s$^{2}$.}
\label{fig:cme_ht}
\end{figure}
The CME website provides various important parameters of CMEs, namely, central angle, angular width, height of CME relative to solar radius, time, mass, and kinetic energy. Using the height (i.e., front of the CME) and time information, the height-time (HT) plot of CME is produced (left panel; Figure~\ref{fig:cme_ht}). Further, this data is fitted using a straight line (blue line) and 2$^{nd}$ order polynomial (red line). Both the functions (i.e., straight line and 2$^{nd}$ order polynomial) seem appropriate to characterize the observation; however, the 2$^{nd}$ order polynomial (red curve) follows the data points more closely than the straight line (blue line). The fitted straight line is as follows.\\

 \begin{equation}
 y = 0.0216~{\times}~t~{+}~3.03  
 \end{equation}

Now, using the slope value (i.e., 0.0216), we find the linear velocity of the CME is around 250 km/s. Further, we estimated the linear velocity at each time with respect to the first time (i.e., 18:24:05~UT), and the linear velocity is displayed with time in the right panel of Figure~\ref{fig:cme_ht}. Initially, the speed decreases from 328 to 280 km/s, while again it increases from 280 to 350 km/s. Later on (after $\sim$100 minutes), the velocity decreases only, and this particular phase is fitted with a straight line (red line). The straight line equation is mentioned at the top of the panel, and using the slope of the straight line, we found that the acceleration is $-$4.21 m/s$^{2}$, i.e., the CME is decelerating after 100 minutes with a deceleration of 4.21 m/s$^2$.  

\section{Summary, Discussion, and Conclusions} \label{sec:disc}
We have performed a detailed investigation of a jet-like prominence eruption using imaging and spectroscopic observations. The main findings are summarized below. 
\begin{itemize}
    \item A prominence is visible before the formation of the blowout jet, and it rises slowly with a speed of $\sim$ 33 km/s, followed by the rapid rise with a speed of $\sim$ 338 km/s. Two weak solar flares (i.e., C4.0 and C4.2) occurred during this event.  
    
    \item Further, it is found that a CME is also associated with this event, and it propagates with a linear speed of 250 km/s. It is also found that initially the CME accelerates while, after $\sim$ 100 minutes, it decelerates with the deceleration of 4.21 m/s$^{2}$.    
      
    \item   During the rapid rise phase, one leg of the prominence is completely broken, which forms a large rotating blowout jet. The blowout jet is made of several plasma threads, and plasma propagates up with speeds ranging from 125 to 594 km/s. Similarly, the downfall speeds range from 43 to 158 km/s for different plasma threads. The lowest downfall speed (i.e., 43 km/s) is the speed of plasma falling from the prominence break point. 

    \item The jet's spire close to the base is expanding at a speed of around 30$-$32 km/s (panels (f) and (g); Figure~\ref{fig:td_across}).  The width of the jet's spire changes from $\sim$15 Mm to $\sim$ 30~Mm.  
    
    \item The TD analysis (across the jet) shows the delay in the transverse motion with distance along the blowout jet. The transverse motions start later at higher heights (panel (a); Figure~\ref{fig:td_across}) than at lower heights (panel (d); Figure~\ref{fig:td_across}), i.e., the transverse oscillations are propagating higher from lower heights. This delay in time is due to the unwinding of the twist (i.e., most probably propagating torsional Alfv{\'e}n waves), and hence, this untwisting front is propagating with a speed of around (i.e., 267 km/s).

    \item The jet plasma column shows large coherent transverse oscillation, see the white dashed path in panel (c) of Figure~\ref{fig:td_across}. Note that not only a single large oscillation, but various plasma threads of the blowout jet show individual oscillations. We have distinctly identified 4 individual decayless oscillations (Figure~\ref{fig:osci}), and they are fitted with a simple sine function. While the larger oscillation decays over time, it is characterized by a combination of sine and exponential functions (with a negative exponent). With the help of these fittings, we have estimated the amplitude, time period, and transverse velocity of these oscillations, which are mentioned in the table~\ref{tab:osci_para}. 
   
    \item We investigate the Doppler velocity structure of the jet, using the Si~{ \sc iv} line, and the Doppler velocity structure is useful to know the rotating/helical motions of solar jets (e.g., \cite{2015ApJ...801...83C, Kayshap2018, 2021MNRAS.505.5311K}). However, despite the existence of a small patch of redshift on the edge of the jet, we found no convincing signature of rotational motion of the jet (see Section~\ref{sect:spectra}). 
    
    \item The Si~{\sc iv} spectral line from the lower spire of the blowout jet is forming in optically thick conditions, while the Si~{\sc iv} ratio from the upper spire of the jet is close to 2, i.e., here the Si~{\sc iv} line is forming in optically thin conditions.
    
    \item The electron density estimation is performed using the O~{\sc iv} lines, and it is found that initially the electron density is increasing with height, and later on the electron density is decreasing. Next, we correlate the line ratio and electron density only from the lower part of the spire, and the line ratio and electron density are positively correlated.  
\end{itemize}
First of all, we have seen the slow rise of the prominence before its eruption, which is a common attribute of the prominence eruption (e.g., \citealt{1980SoPh...65..357T, 1988ApJ...328..824K, 2007A&A...472..967C, 2008SoPh..247..321S, 2009SoPh..256...57L}). Further, one leg of the prominence breaks, and this type of breaking of one leg of the prominence is known as asymmetric prominence eruption (e.g., \citealt{2006A&A...458..965C, 2009ApJ...691.1079L, 2010ApJ...721L.193L}). Various models have been developed for prominence eruptions, for example, models based on the instabilities in the magnetic flux rope (e.g., \citealt{2005ApJ...630L..97T}), magnetic breakout model (\citealt{1999ApJ...510..485A}), and tether-cutting model (e.g., \citealt{2001ApJ...552..833M, 2006GMS...165...43M}). In the first type of models, the helical kink/torus instabilities trigger the prominence eruption (e.g., \citealt{2005ApJ...630L..97T, 2012NatCo...3..747Z}). While the other two models (i.e., magnetic breakout and tether cutting models) are based on the magnetic reconnection (e.g., \citealt{2001ApJ...552..833M, 2006GMS...165...43M}). Unfortunately, the source region of this prominence is located on the other side of the Sun; therefore, we don't have access to the magnetic configuration of this event. Hence, it would be hard which eruption model best explain the event. Nevertheless, some important findings revealed by multiwavelength analysis reveals should be mentioned. Prior to the breaking of one leg of the prominence, significant brightenings occur all over the prominence in the cool temperature filters (i.e., AIA~304~{\AA} and IRIS/SJI~2796~{\AA}). Simultaneously, the SXR starts to increase from 17:48~UT, i.e., initiation of the C-class solar flare. Next, it is important to note that the maximum of the SXR occurs (i.e., the maximum of the C-class solar flare) when the northern leg of the prominence breaks. In hot temperature filter AIA~94~{\AA}, the small-scale brightenings exist just below the prominence before the asymmetric eruption (see panels (b) and (c); Figure~\ref{fig:aia94}). Later, the brightness is spread all over the base region, and the shape of the prominence is lost.\\ 


After the eruption, the bipolar shape of prominence is converted into a plasma column, which appears as the blowout type of jet as proposed by \cite{2010ApJ...720..757M}. The solar jets are divided into two categories: standard jets and blowout jets (\cite{2010ApJ...720..757M}). The standard jets form due to the magnetic reconnection between two opposite polarities, mainly, a closed magnetic bipolar field and slanted/horizontal open fields (e.g., \citealt{Shibata1992, Yokoyama1995, Yokoyama1996, 2004ApJ...614.1042M,  Shibata2007, Nisizuka2008, Kayshap2013a}). However, not the magnetic reconnection, but the confined filament eruptions make most of the standard jets, as recently reported by \cite{2022ApJ...933...12M}. In the case of the blowout jets, the sheared core field emerged, and during the reconnection, the core field erupts to form the blowout jet (e.g., \citealt{2010ApJ...720..757M, 2013ApJ...769..134M, 2013ApJ...769L..21A, 2016A&A...594A..64P}). In a notable work, \cite{Kayshap2013b} reported the internal reconnection within the omge-shaped flux rope, which breaks the upper part of the flux rope and leads to the formation of a blowout jet at the north polar region. Later on, the formation of a blowout jet is reported via the eruption of flux rope/mini-filament in various works (e.g., \citealt{2012ApJ...745..164S, 2014ApJ...783...11A, 2018MNRAS.476.1286J, 2015Natur.523..437S, 2016A&A...594A..64P, 2017ApJ...851...67S, 2017ApJ...835...35H, 2019ApJ...883..104S, 2024SoPh..299...88K}). Hence, the eruption of a shared magnetic flux rope/mini-filament is an important element of solar blowout jets (for more details, see recent review article by \citealt{2021RSPSA.47700217S, 2023SSRv..219...33G}). In the present event, we are witnessing the slow rise of prominence following the weakening and breaking of the north leg of the prominence. Further, the untwisting motion of prominence plasma occurs, and it appears as a blowout jet. This scenario is somewhat similar to the breaking of a small-scale flux rope and formation of a blowout jet as reported by \cite{Kayshap2013b}, but at a bigger scale. Here, it is also worth mentioning that \cite{2015Natur.523..437S} reported the formation of jets due to the eruption of the mini-filaments. Hence, we emphasize that not only small-scale flux-rope/mini-filament triggers the blowout jet, but the usual bigger prominence can also appear as a blowout jet after the asymmetric eruption (i.e., eruption of one leg). Our study is on the kinematics and plasma diagnostics of this blowout jet.\\

{\it \textbf{Motions Along the Jet:}} In the initial phase, the prominence ascends slowly with a speed of around 33 km/s, and later on, the prominence rises rapidly with a speed of 338 km/s prior to breaking of the north leg of prominence, i.e., formation of blowout jet. This blowout jet is composed with different plasma threads. Next, the different plasma threads of the blowout jet propagate upwards at different speeds, i.e., $\sim$125 to 594 km/s. On the other hand, the downflow speed varies from around 43 to 157 km/s, only.\\ 

{\it \textbf{Axial Motions:}} The untwisting motion (i.e., unwinding the twist) can be explained with the help of the observed delay of the transverse motion with distance along the jet. In the current observations, the time-delay in the occurrence of transverse oscillation at different heights (i.e, cuts), and it justifies that these oscillations propagate along the jet at higher altitudes (see Figure~\ref{fig:td_across}). We located one clear trough and tracked it with height, i.e., we estimated the position and time of this trough at different altitudes along the jet (see the white asterisk from panels (a) to (d); Figure~\ref{fig:td_across}). With the help of this information, the phase speed is estimated, and it comes to 267 km/s. This phase speed corresponds to the untwisting front (\cite{1985PASJ...37...31S}), and should propagate with the Alfv{\'e}n speeds. The phase speed in the current work is significantly lower than the phase speed values reported by \cite{2009ApJ...707L..37L} for a chromospheric jet, i.e., 744$\pm$11 km/s for the first crest and 348$\pm$5 km/s in the second crest.\\ 

{\it \textbf{Transverse Motion:}} The blowout jet shows the oscillatory transverse motion of the entire bundle of plasma in the anticlockwise direction. This transverse oscillatory motion is due to the unwinding motion of the twisted magnetic field. Various models predict that the solar jets are composed of twisted (helical) field lines, and later these helical field lines can undergo the untwisting motion (e.g., \citealt{1985PASJ...37...31S, 1986SoPh..103..299S, 1996ApJ...464.1016C, 2011ApJ...735L..43S, 2004ApJ...610.1129J, 2014ApJ...789L..19F, 2021MNRAS.505.5311K}). The projection of rotations about an axis in the sky plane appears as oscillatory transverse motion (e.g.,\citealt{2009ApJ...707L..37L}). The entire bundle of blowout jet plasma oscillates with a transverse velocity of 126.18$pm$7.27 km/s (main oscillation; Table~\ref{tab:osci_para}). Here, it is important to note that deduced transverse velocity in the present observation is comparable with previously reported values, for instances, whip-like/transverse motions with 120 km/s in surges/jets (e.g., \citealt{1996ApJ...464.1016C, 2007PASJ...59S.745S}) and the transverse velocity of around 151$\pm$6 km/s in chromospheric jet (\citealt{2009ApJ...707L..37L}).\\
Solar jets can have small-scale threads (see Figure 1 of \citealt{2009ApJ...707L..37L}), and they individually undergo twisting motions. Not only solar jets, but the prominences can also have various plasma threads and exhibit transverse motions (e.g., \citealt{2007Sci...318.1577O, 2015ApJ...814L..17S, 2017ApJ...851...47Z}). Here, in the present observation, the blowout has different plasma threads and oscillates transversely at different speeds. We have investigated the transverse oscillations in four distinct plasma threads, and the transverse velocity varies between 30 and 65 km/s with the period ranging from 406{--}811 s(see Table~\ref {tab:osci_para}). The reported values are consistent with the previously reported values (e.g., \citealt{2009ApJ...707L..37L}).\\ 

{\it \textbf{Line Ratio \& Electron Density:}} The Si~{\sc iv} line ratio is estimated in the jet region, and it tells under which conditions (i.e., optically thick or thin) these lines are forming. It is found near the base of the blowout jet; the lines are forming in optically thick conditions, while the profiles from the spire are forming in optically thin conditions. Next, with the help O~{\sc iv} intensity ratio and CHIANTI atomic database, the electron density is estimated along the jet. And, it is found that the electron density increases till the height of 7.8 Mm from the base of the blowout jet, and further it decreases.it decrease. It means that TR spectral lines are forming in the optically thick conditions in the higher electron density regions.\\

The main identification of the solar jets is a well-collimated flow of plasma along the magnetic field lines. However, this is not always the case. \citealt{2010ApJ...720..757M} has shown the existence of a blowout jet: not well collimated but having a wider spire than the standard jets. As per the scientific schematic provided by \cite{2010ApJ...720..757M}, the multiple reconnection sites are responsible for the formation of a wider spire of the jet. Further, it has shown that the eruption of the mini-filament/prominence at the base triggers blowout jets (e.g., \citealt{2013ApJ...769..134M, 2015Natur.523..437S, 2024ApJ...977..141K}). This scenario is similar to the prominence/filament eruption in coronal mass ejections (CMEs; \citealt{2010ApJ...720..757M}). But instead of a mini-filament/prominence, if a large prominence erupts, then we don't see a well collimated eruption, rather a widespread eruption of an irregular shape (see \citealt{2014LRSP...11....1P}). If the size of filament is small (i.e., mini-filament), then it probably triggers jet-like structures (e.g., \citealt{2010ApJ...720..757M, 2013ApJ...769..134M, 2015Natur.523..437S,2024ApJ...977..141K}), and in some cases, this can produce the jet-like CMEs (e.g., \citealt{2020ApJ...901...94J}). Here, we believe the size of the erupting prominence would be important to judge whether the eruption will appear as a solar jet or prominence eruption. 

In the present observation, the size of erupting prominence is comparable to the size of erupting mini-filaments reported previously, which form the solar jets (e.g., \citealt{2012ApJ...745..164S, 2016ApJ...832L...7P}). Therefore, the eruption in the present work appears as a blowout-type jet. This work clearly shows the breaking of prominence, which is hard to see in on-disk solar jets. Previously, \cite{1999ApJ...520L..71W} has reported the jet-like prominence in the He~{\sc ii} 304~{\AA}. Just after breaking one leg, the prominence appears as a blowout type of jet, i.e., a broad plasma column. While later on, the spire of blowout jet is significantly bent from one location, and the spire above the bend becomes very wide (i.e., around 60 Mm). And, it seems that the latter part of the event appears as the eruption of prominence. This particular event is an important landmark as it shifts from the jet-like structure towards the highly bended wide spire, i.e., closer towards the prominence eruption. Hence, we believe this particular event is a hybrid of a solar jet and a prominence eruption.




\begin{acknowledgments}
The work in this article is part of the project DynaSun, which has received funding under the Horizon Europe programme of the European Union under grant agreement No. 101131534. Views and opinions expressed are, however, those of the author(s) only and do not necessarily reflect those of the European Union; therefore, the European Union cannot be held responsible for them. Y. S. was supported by the Natural Science Foundation of China (12573059, 12173083) and the Specialized Research Fund for the State Key Laboratory of Solar Activity and Space Weather. IRIS is a NASA small explorer mission developed and operated by LMSAL, with mission operations executed at NASA Ames Research Center and major contributions to downlink communications funded by ESA and the Norwegian Space Centre. We also acknowledge the use of imaging observations provided by AIA/SDO.   
\end{acknowledgments}

\begin{contribution}
The data analysis is done by PK, AKB, and SB. The science case is developed by PK, YS, and PJ. The first draft of the manuscript was prepared by PK and SB, and the suggestions were given by YS and PJ. And, the suggestions are implemented by PK, PJ, and SB to produce the final manuscript.   



\end{contribution}








\begin{thebibliography}{}

\bibitem[M. Adams et al.(2014)]{2014ApJ...783...11A}
Adams, M., Sterling, A. C., Moore, R. L., \& Gary, G. A.\ 2014, \apj, 783, 11.
\newblock \href{https://doi.org/10.1088/0004-637X/783/1/11}{\path{doi:10.1088/0004-637X/783/1/11}}

\bibitem[S. K. Antiochos et al.(1999)]{1999ApJ...510..485A}
Antiochos, S. K., DeVore, C. R., \& Klimchuk, J. A.\ 1999, \apj, 510, 485.
\newblock \href{https://doi.org/10.1086/306563}{\path{doi:10.1086/306563}}

\bibitem[V. Archontis \& A. W. Hood(2013)]{2013ApJ...769L..21A}
Archontis, V. \& Hood, A.~W.\ 2013, \apjl, 769, L21.
\newblock \href{https://doi.org/10.1088/2041-8205/769/2/L21}{\path{doi:10.1088/2041-8205/769/2/L21}}

\bibitem[B. S. Babu et al.(2024a)]{2024Ap&SS.369...61B}
Babu, B. S., Kayshap, P., \& Tripathi, S. C.\ 2024, \apss, 369, 61.
\newblock \href{https://doi.org/10.1007/s10509-024-04323-5}{\path{doi:10.1007/s10509-024-04323-5}}

\bibitem[B. S. Babu et al.(2024b)]{2024MNRAS.528.2474B}
Babu, B. S., Kayshap, P., Tripathi, S. C., Jel{\'\i}nek, P., \& Dwivedi, B. N.\ 2024, \mnras, 528, 2474. \newblock \href{https://doi.org/10.1093/mnras/stae166}{\path{doi:10.1093/mnras/stae166}}

\bibitem[B. S. Babu et al.(2025)]{2025Ap&SS.370...13B}
Babu, B. S., Kayshap, P., \& Tripathi, S. C.\ 2025, \apss, 370, 13.
\newblock \href{https://doi.org/10.1007/s10509-025-04404-z}{\path{doi:10.1007/s10509-025-04404-z}}

\bibitem[G. E. Brueckner et al.(1995)]{1995SoPh..162..357B}
Brueckner, G. E., et. al, 1995, \solphys, 162, 357. \newblock \href{https://doi.org/10.1007/BF00733434}{\path{doi:10.1007/BF00733434}}

\bibitem[R. C. Canfield et al.(1996)]{1996ApJ...464.1016C}
Canfield, R. C., Reardon, K. P., Leka, K. D., Shibata, K., Yokoyama, T., \& Shimojo, M.\ 1996, \apj, 464, 1016.
\newblock \href{https://doi.org/10.1086/177389}{\path{doi:10.1086/177389}}

\bibitem[M. C. M. Cheung et al.(2015)]{2015ApJ...801...83C}
Cheung, M. C. M., et. al, 2015, \apj, 801, 83.
\newblock \href{https://doi.org/10.1088/0004-637X/801/2/83}{\path{doi:10.1088/0004-637X/801/2/83}}

\bibitem[C. Chifor et al.(2006)]{2006A&A...458..965C}
Chifor, C., Mason, H. E., Tripathi, D., Isobe, H., \& Asai, A.\ 2006, \aap, 458, 965.
\newblock \href{https://doi.org/10.1051/0004-6361:20065687}{\path{doi:10.1051/0004-6361:20065687}}

\bibitem[C. Chifor et al.(2007)]{2007A&A...472..967C}
Chifor, C., Tripathi, D., Mason, H. E., \& Dennis, B. R.\ 2007, \aap, 472, 967.
\newblock \href{https://doi.org/10.1051/0004-6361:20077771}{\path{doi:10.1051/0004-6361:20077771}}

\bibitem[B. De Pontieu et al.(2014)]{2014SoPh..289.2733D}
De Pontieu, B., et. al, 2014, \solphys, 289, 2733.
\newblock \href{https://doi.org/10.1007/s11207-014-0485-y}{\path{doi:10.1007/s11207-014-0485-y}}

\bibitem[G. Del Zanna \& H. E. Mason(2018)]{2018LRSP...15....5D}
Del Zanna, G., \& Mason, H. E.\ 2018, Living Reviews in Solar Physics, 15, 5.
\newblock \href{https://doi.org/10.1007/s41116-018-0015-3}{\path{doi:10.1007/s41116-018-0015-3}}

\bibitem[F. Fang et al.(2014)]{2014ApJ...789L..19F}
Fang, F., Fan, Y., \& McIntosh, S. W.\ 2014, \apjl, 789, L19.
\newblock \href{https://doi.org/10.1088/2041-8205/789/1/L19}{\path{doi:10.1088/2041-8205/789/1/L19}}

\bibitem[S. E. Gibson(2018)]{2018LRSP...15....7G}
Gibson, S. E.\ 2018, Living Reviews in Solar Physics, 15, 7.
\newblock \href{https://doi.org/10.1007/s41116-018-0016-2}{\path{doi:10.1007/s41116-018-0016-2}}

\bibitem[H. R. Gilbert et al.(2000)]{2000ApJ...537..503G}
Gilbert, H. R., Holzer, T. E., Burkepile, J. T., \& Hundhausen, A. J.\ 2000, \apj, 537, 503.
\newblock \href{https://doi.org/10.1086/309030}{\path{doi:10.1086/309030}}

\bibitem[N. Gopalswamy et al.(2003)]{2003ApJ...586..562G}
Gopalswamy, N., Shimojo, M., Lu, W., Yashiro, S., Shibasaki, K., \& Howard, R. A.\ 2003, \apj, 586, 562.
\newblock \href{https://doi.org/10.1086/367614}{\path{doi:10.1086/367614}}

\bibitem[S. Gosain et al.(2009)]{2009SoPh..259...13G}
Gosain, S., Schmieder, B., Venkatakrishnan, P., Chandra, R., \& Artzner, G.\ 2009, \solphys, 259, 13. \newblock \href{https://doi.org/10.1007/s11207-009-9448-0}{\path{doi:10.1007/s11207-009-9448-0}}

\bibitem[S. Gun{\'a}r et al.(2023)]{2023SSRv..219...33G}
Gun{\'a}r, S., Labrosse, N., Luna, M., Schmieder, B., Heinzel, P., Kucera, T. A., Levens, P. J., L{\'o}pez Ariste, A., Mackay, D. H., \& Zapi{\'o}r, M.\ 2023, \ssr, 219, 33. \newblock \href{https://doi.org/10.1007/s11214-023-00976-w}{\path{doi:10.1007/s11214-023-00976-w}}

\bibitem[Y. Guo et al.(2024)]{2024ApJ...970..110G}
Guo, Y., Hou, Y., Li, T., Shen, Y., Wang, J., Zhang, J., Zheng, J., Wang, D., \& Mei, L.\ 2024, \apj, 970, 110. \newblock \href{https://doi.org/10.3847/1538-4357/ad54b8}{\path{doi:10.3847/1538-4357/ad54b8}}

\bibitem[J. Hong et al.(2017)]{2017ApJ...835...35H}
Hong, J., Jiang, Y., Yang, J., Li, H., \& Xu, Z.\ 2017, \apj, 835, 35.
\newblock \href{https://doi.org/10.3847/1538-4357/835/1/35}{\path{doi:10.3847/1538-4357/835/1/35}}

\bibitem[K. Hori \& J. L. Culhane(2002)]{2002A&A...382..666H}
Hori, K., \& Culhane, J. L.\ 2002, \aap, 382, 666. \newblock \href{https://doi.org/10.1051/0004-6361:20011658}{\path{doi:10.1051/0004-6361:20011658}}

\bibitem[R. Hosseini et al.(2024)]{2024MNRAS.529.3424H}
Hosseini, R., Kayshap, P., Alipour, N., \& Safari, H.\ 2024, \mnras, 529, 3424.
\newblock \href{https://doi.org/10.1093/mnras/stae356}{\path{doi:10.1093/mnras/stae356}}

\bibitem[D. E. Innes et al.(1997)]{Innes1997}
Innes, D. E., Inhester, B., Axford, W. I., \& Wilhelm, K.\ 1997, \nat, 386, 811.
\newblock \href{https://doi.org/10.1038/386811a0}{\path{doi:10.1038/386811a0}}

\bibitem[H. Isobe \& D. Tripathi(2006)]{2006A&A...449L..17I}
Isobe, H. \& Tripathi, D.\ 2006, \aap, 449, L17.
\newblock \href{https://doi.org/10.1051/0004-6361:20064942}{\path{doi:10.1051/0004-6361:20064942}}

\bibitem[P. Jelinek \& M. Karlicky(2026)]{Jelnek2026}
Jelinek, P., \& Karlicky, M.\ 2026, \aap, submitted.

\bibitem[P. Jelinek et al.(2015)]{Jelinek2015}
Jelinek, P., Srivastava, A. K., Murawski, K., Kayshap, P., \& Dwivedi, B. N.\ 2015, \aap, 581, A131. \newblock \href{https://doi.org/10.1051/0004-6361/201424234}{\path{doi:10.1051/0004-6361/201424234}}

\bibitem[P. Jibben \& R. C. Canfield(2004)]{2004ApJ...610.1129J}
Jibben, P., \& Canfield, R. C.\ 2004, \apj, 610, 1129.
\newblock \href{https://doi.org/10.1086/421727}{\path{doi:10.1086/421727}}

\bibitem[N. C. Joshi et al.(2018)]{2018MNRAS.476.1286J}
Joshi, N. C., Nishizuka, N., Filippov, B., Magara, T., \& Tlatov, A. G.\ 2018, \mnras, 476, 1286. \newblock \href{https://doi.org/10.1093/mnras/sty322}{\path{doi:10.1093/mnras/sty322}}

\bibitem[R. Joshi et al.(2020)]{2020ApJ...901...94J}
Joshi, R., Wang, Y., Chandra, R., Zhang, Q., Liu, L., \& Li, X.\ 2020, \apj, 901, 94. \newblock \href{https://doi.org/10.3847/1538-4357/abaf5a}{\path{doi:10.3847/1538-4357/abaf5a}}

\bibitem[S. W. Kahler et al.(1988)]{1988ApJ...328..824K}
Kahler, S. W., Moore, R. L., Kane, S. R., \& Zirin, H.\ 1988, \apj, 328, 824.
\newblock \href{https://doi.org/10.1086/166340}{\path{doi:10.1086/166340}}

\bibitem[P. Kayshap et al.(2013a)]{Kayshap2013a}
Kayshap, P., Srivastava, A.~K., \& Murawski, K.\ 2013, \apj, 763, 24.
\newblock \href{https://doi.org/10.1088/0004-637X/763/1/24}{\path{doi:10.1088/0004-637X/763/1/24}}

\bibitem[P. Kayshap et al.(2013)]{Kayshap2013b}
Kayshap, P., Srivastava, A.~K., Murawski, K., \& Tripathi, D.\ 2013, \apjl, 770, L3.
\newblock \href{https://doi.org/10.1088/2041-8205/770/1/L3}{\path{doi:10.1088/2041-8205/770/1/L3}}

\bibitem[P. Kayshap et al.(2015)]{Kayshap2015}
Kayshap, P., Banerjee, D., \& Srivastava, A.~K. 2015, Solar Physics, 290, 2889.
\newblock \href{https://doi.org/10.1007/s11207-015-0763-3}{\path{doi:10.1007/s11207-015-0763-3}}

\bibitem[P. Kayshap et al.(2015)]{2015SoPh..290.2889K}
Kayshap, P., Banerjee, D., \& Srivastava, A.~K.\ 2015, \solphys, 290, 2889.
\newblock \href{https://doi.org/10.1007/s11207-015-0763-3}{\path{doi:10.1007/s11207-015-0763-3}}

\bibitem[P. Kayshap \& B. N. Dwivedi(2017)]{2017SoPh..292..108K}
Kayshap, P. \& Dwivedi, B.~N.\ 2017, \solphys, 292, 108.
\newblock \href{https://doi.org/10.1007/s11207-017-1132-1}{\path{doi:10.1007/s11207-017-1132-1}}

\bibitem[P. Kayshap et al.(2018)]{Kayshap2018}
Kayshap, P., Murawski, K., Srivastava, A.~K., \& Dwivedi, B.~N.\ 2018a, \aap, 616, A99. \newblock \href{https://doi.org/10.1051/0004-6361/201730990}{\path{doi:10.1051/0004-6361/201730990}}

\bibitem[P. Kayshap et al.(2018)]{2018ApJ...864...21K}
Kayshap, P., Tripathi, D., Solanki, S.~K., \& Peter, H.\ 2018b, \apj, 864, 21.
\newblock \href{https://doi.org/10.3847/1538-4357/aad2d9}{\path{doi:10.3847/1538-4357/aad2d9}}

\bibitem[P. Kayshap et al.(2021)]{2021MNRAS.505.5311K}
Kayshap, P., Singh Payal, R., Tripathi, S.~C., \& Padhy, H.\ 2021a, \mnras, 505, 5311. \newblock \href{https://doi.org/10.1093/mnras/stab1663}{\path{doi:10.1093/mnras/stab1663}}

\bibitem[P. Kayshap et al.(2021)]{2021ApJ...906..121K}
Kayshap, P., Tripathi, D., \& Jel{\'\i}nek, P.\ 2021b, \apj, 906, 121.
\newblock \href{https://doi.org/10.3847/1538-4357/abcc6f}{\path{doi:10.3847/1538-4357/abcc6f}}

\bibitem[P. Kayshap \& P. R. Young(2023)]{2023MNRAS.526..383K}
Kayshap, P. \& Young, P.~R.\ 2023, \mnras, 526, 383. \newblock \href{https://doi.org/10.1093/mnras/stad2761}{\path{doi:10.1093/mnras/stad2761}}

\bibitem[P. Kayshap et al.(2024)]{2024SoPh..299...88K}
Kayshap, P., Karpen, J.~T., \& Kumar, P.\ 2024, \solphys, 299, 88.
\newblock \href{https://doi.org/10.1007/s11207-024-02315-w}{\path{doi:10.1007/s11207-024-02315-w}}

\bibitem[P. Kayshap \& Young(2024)]{2024ApJ...977..141K}
Kayshap, P. \& Young, P.~R.\ 2024, \apj, 977, 141. \newblock \href{https://doi.org/10.3847/1538-4357/ad901d}{\path{doi:10.3847/1538-4357/ad901d}}

\bibitem[T. A. Kucera \& E. Landi(2008)]{2008ApJ...673..611K}
Kucera, T.~A. \& Landi, E.\ 2008, \apj, 673, 611.
\newblock \href{https://doi.org/10.1086/523694}{\path{doi:10.1086/523694}}

\bibitem[P. Kumar et al.(2018)]{2018ApJ...854..155K}
Kumar, P., Karpen, J.~T., Antiochos, S.~K., Wyper, P.~F., DeVore, C.~R., \& DeForest, C.~E.\ 2018, \apj, 854, 155. \newblock \href{https://doi.org/10.3847/1538-4357/aaab4f}{\path{doi:10.3847/1538-4357/aaab4f}}

\bibitem[Ku{\'z}ma et al.(2017)]{2017ApJ...849...78K}
Ku{\'z}ma, B{\l}a{\.z}ej, Murawski, K., Kayshap, P., W{\'o}jcik, D., Srivastava, A.~K., \& Dwivedi, B.~N.\ 2017, \apj, 849, 78. \newblock \href{https://doi.org/10.3847/1538-4357/aa8ea1}{\path{doi:10.3847/1538-4357/aa8ea1}}

\bibitem[N. Labrosse et al.(2010)]{2010SSRv..151..243L}
Labrosse, N., Heinzel, P., Vial, J.-C., Kucera, T., Parenti, S., Gun{\'a}r, S., Schmieder, B., \& Kilper, G.\ 2010, \ssr, 151, 243.
\newblock \href{https://doi.org/10.1007/s11214-010-9630-6}{\path{doi:10.1007/s11214-010-9630-6}}

\bibitem[J. R. Lemen et al.(2012)]{2012SoPh..275...17L}
Lemen, J.~R., Title, A.~M., Akin, D.~J., et al.\ 2012, \solphys, 275, 17.
\newblock \href{https://doi.org/10.1007/s11207-011-9776-8}{\path{doi:10.1007/s11207-011-9776-8}}

\bibitem[D. Li et al.(2018)]{2018ApJ...863..192L}
Li, D., Shen, Y., Ning, Z., Zhang, Q., \& Zhou, T.\ 2018, \apj, 863, 192.
\newblock \href{https://doi.org/10.3847/1538-4357/aad33f}{\path{doi:10.3847/1538-4357/aad33f}}

\bibitem[X. Li et al.(2012)]{2012ApJ...752L..22L}
Li, X., Morgan, H., Leonard, D., \& Jeska, L.\ 2012, \apjl, 752, L22.
\newblock \href{https://doi.org/10.1088/2041-8205/752/2/L22}{\path{doi:10.1088/2041-8205/752/2/L22}}

\bibitem[P. C. Liewer et al.(2009)]{2009SoPh..256...57L}
Liewer, P.~C., De Jong, E.~M., Hall, J.~R., Howard, R.~A., Thompson, W.~T., Culhane, J.~L., Bone, L., \& van Driel-Gesztelyi, L.\ 2009, \solphys, 256, 57.
\newblock \href{https://doi.org/10.1007/s11207-009-9363-4}{\path{doi:10.1007/s11207-009-9363-4}}

\bibitem[Y. Lin(2011)]{2011SSRv..158..237L}
Lin, Y.\ 2011, \ssr, 158, 237. \newblock \href{https://doi.org/10.1007/s11214-010-9672-9}{\path{doi:10.1007/s11214-010-9672-9}}

\bibitem[C. Liu et al.(2010)]{2010ApJ...721L.193L}
Liu, C., Lee, J., Jing, J., Liu, R., Deng, N., \& Wang, H.\ 2010, \apjl, 721, L193.
\newblock \href{https://doi.org/10.1088/2041-8205/721/2/L193}{\path{doi:10.1088/2041-8205/721/2/L193}}

\bibitem[R. Liu et al.(2009)]{2009ApJ...691.1079L}
Liu, R., Alexander, D., \& Gilbert, H.~R.\ 2009, \apj, 691, 1079.
\newblock \href{https://doi.org/10.1088/0004-637X/691/2/1079}{\path{doi:10.1088/0004-637X/691/2/1079}}

\bibitem[W. Liu et al.(2009)]{2009ApJ...707L..37L}
Liu, W., Berger, T.~E., Title, A.~M., \& Tarbell, T.~D.\ 2009, \apjl, 707, L37.
\newblock \href{https://doi.org/10.1088/0004-637X/707/1/L37}{\path{doi:10.1088/0004-637X/707/1/L37}}

\bibitem[S. K. Mishra et al.(2023)]{Mishra2023}
Mishra, S.~K., Sangal, K., Kayshap, P., Jel{\'\i}nek, P., Srivastava, A.~K., \& Rajaguru, S.~P.\ 2023, \apj, 945, 113. \newblock \href{https://doi.org/10.3847/1538-4357/acb058}{\path{doi:10.3847/1538-4357/acb058}}

\bibitem[T. Miyagoshi \& T. Yokoyama(2004)]{2004ApJ...614.1042M}
Miyagoshi, T. \& Yokoyama, T.\ 2004, \apj, 614, 1042. \newblock \href{https://doi.org/10.1086/423731}{\path{doi:10.1086/423731}}

\bibitem[R. L. Moore et al.(2001)]{2001ApJ...552..833M}
Moore, R.~L., Sterling, A.~C., Hudson, H.~S., \& Lemen, J.~R.\ 2001, \apj, 552, 833.
\newblock \href{https://doi.org/10.1086/320559}{\path{doi:10.1086/320559}}

\bibitem[R. L. Moore et al.(2010)]{2010ApJ...720..757M}
Moore, R.~L., Cirtain, J.~W., Sterling, A.~C., \& Falconer, D.~A.\ 2010, \apj, 720, 757. \newblock \href{https://doi.org/10.1088/0004-637X/720/1/757}{\path{doi:10.1088/0004-637X/720/1/757}}

\bibitem[R. L. Moore et al.(2013)]{2013ApJ...769..134M}
Moore, R.~L., Sterling, A.~C., Falconer, D.~A., \& Robe, D.\ 2013, \apj, 769, 134.
\newblock \href{https://doi.org/10.1088/0004-637X/769/2/134}{\path{doi:10.1088/0004-637X/769/2/134}}

\bibitem[R. L. Moore \& A. C. Sterling(2006)]{2006GMS...165...43M}
Moore, R.~L. \& Sterling, A.~C.\ 2006, Geophysical Monograph Series, 165, 43.
\newblock \href{https://doi.org/10.1029/9781118666203.ch6}{\path{doi:10.1029/9781118666203.ch6}}

\bibitem[R. L. Moore et al.(2022)]{2022ApJ...933...12M}
Moore, R.~L., Panesar, N.~K., Sterling, A.~C., \& Tiwari, S.~K.\ 2022, \apj, 933, 12. \newblock \href{https://doi.org/10.3847/1538-4357/ac6181}{\path{doi:10.3847/1538-4357/ac6181}}

\bibitem[R. H. Munro et al.(1979)]{1979SoPh...61..201M}
Munro, R.~H., Gosling, J.~T., Hildner, E., MacQueen, R.~M., Poland, A.~I., \& Ross, C.~L.\ 1979, \solphys, 61, 201.
\newblock \href{https://doi.org/10.1007/BF00155456}{\path{doi:10.1007/BF00155456}}

\bibitem[N. Nishizuka et al.(2008)]{Nisizuka2008}
Nishizuka, N., Shimizu, M., Nakamura, T., Otsuji, K., Okamoto, T.~J., Katsukawa, Y., \& Shibata, K.\ 2008, \apjl, 683, L83.
\newblock \href{https://doi.org/10.1086/591445}{\path{doi:10.1086/591445}}

\bibitem[L. Ofman et al.(1998)]{1998SoPh..183...97O}
Ofman, L., Kucera, T.~A., Mouradian, Z., \& Poland, A.~I.\ 1998, \solphys, 183, 97.
\newblock \href{https://doi.org/10.1023/A:1005052923972}{\path{doi:10.1023/A:1005052923972}}

\bibitem[T. J. Okamoto et al.(2007)]{2007Sci...318.1577O}
Okamoto, T.~J., Tsuneta, S., Berger, T.~E., et al.\ 2007, Science, 318, 1577.
\newblock \href{https://doi.org/10.1126/science.1145447}{\path{doi:10.1126/science.1145447}}

\bibitem[N. K. Panesar et al.(2016)]{2016ApJ...832L...7P}
Panesar, N.~K., Sterling, A.~C., Moore, R.~L., \& Chakrapani, P.\ 2016, \apjl, 832, L7. \newblock \href{https://doi.org/10.3847/2041-8205/832/1/L7}{\path{doi:10.3847/2041-8205/832/1/L7}}

\bibitem[V. Pant et al.(2018)]{2018ApJ...860...80P}
Pant, V., Datta, A., Banerjee, D., Chandrashekhar, K., \& Ray, S.\ 2018, \apj, 860, 80. \newblock \href{https://doi.org/10.3847/1538-4357/aac2ba}{\path{doi:10.3847/1538-4357/aac2ba}}

\bibitem[S. Parenti(2014)]{2014LRSP...11....1P}
Parenti, S.\ 2014, Living Reviews in Solar Physics, 11, 1.
\newblock \href{https://doi.org/10.12942/lrsp-2014-1}{\path{doi:10.12942/lrsp-2014-1}}

\bibitem[E. Pariat et al.(2009)]{2009ApJ...691...61P}
Pariat, E., Antiochos, S.~K., \& DeVore, C.~R.\ 2009, \apj, 691, 61.
\newblock \href{https://doi.org/10.1088/0004-637X/691/1/61}{\path{doi:10.1088/0004-637X/691/1/61}}

\bibitem[W. D. Pesnell et al.(2012)]{2012SoPh..275....3P}
Pesnell, W.~D., Thompson, B.~J., \& Chamberlin, P.~C.\ 2012, \solphys, 275, 3.
\newblock \href{https://doi.org/10.1007/s11207-011-9841-3}{\path{doi:10.1007/s11207-011-9841-3}}

\bibitem[H. Peter \& P. G. Judge(1999)]{1999ApJ...522.1148P}
Peter, H. \& Judge, P.~G.\ 1999, \apj, 522, 1148.
\newblock \href{https://doi.org/10.1086/307672}{\path{doi:10.1086/307672}}

\bibitem[V. Polito et al.(2016)]{2016A&A...594A..64P}
Polito, V., Del Zanna, G., Dud{\'\i}k, J., Mason, H.~E., Giunta, A., \& Reeves, K.~K.\ 2016, \aap, 594, A64.
\newblock \href{https://doi.org/10.1051/0004-6361/201628965}{\path{doi:10.1051/0004-6361/201628965}}

\bibitem[Y. K. Rao et al.(2019)]{2019ApJ...874...56R}
Rao, Y.~K., Srivastava, A.~K., Kayshap, P., Wilhelm, K., \& Dwivedi, B.~N.\ 2019, \apj, 874, 56.
\newblock \href{https://doi.org/10.3847/1538-4357/ab06f5}{\path{doi:10.3847/1538-4357/ab06f5}}

\bibitem[B. Schmieder et al.(2008)]{2008SoPh..247..321S}
Schmieder, B., Bommier, V., Kitai, R., Matsumoto, T., Ishii, T.~T., Hagino, M., Li, H., \& Golub, L.\ 2008, \solphys, 247, 321.
\newblock \href{https://doi.org/10.1007/s11207-007-9100-9}{\path{doi:10.1007/s11207-007-9100-9}}

\bibitem[B. Seo et al.(2025)]{2025ApJ...994...47S}
Seo, B., Avila, M., Kayshap, P., Yoo, J., Clevenger, T., Ma, X., Barjatya, A., \& Kim, D.\ 2025, \apj, 994, 47.
\newblock \href{https://doi.org/10.3847/1538-4357/ae119f}{\path{doi:10.3847/1538-4357/ae119f}}

\bibitem[Y. Shen(2021)]{2021RSPSA.47700217S}
Shen, Y.\ 2021, Proceedings of the Royal Society of London Series A, 477, 217.
\newblock \href{https://doi.org/10.1098/rspa.2020.0217}{\path{doi:10.1098/rspa.2020.0217}}

\bibitem[Y. Shen et al.(2011)]{2011ApJ...735L..43S}
Shen, Y., Liu, Y., Su, J., \& Ibrahim, A.\ 2011, \apjl, 735, L43.
\newblock \href{https://doi.org/10.1088/2041-8205/735/2/L43}{\path{doi:10.1088/2041-8205/735/2/L43}}

\bibitem[Y -D. Shen et al.(2011)]{2011RAA....11..594S}
Shen, Y-D., Liu, Y., \& Liu, R.\ 2011, Research in Astronomy and Astrophysics, 11, 594. \newblock \href{https://doi.org/10.1088/1674-4527/11/5/009}{\path{doi:10.1088/1674-4527/11/5/009}}

\bibitem[Y. Shen et al.(2012)]{2012ApJ...745..164S}
Shen, Y., Liu, Y., Su, J., \& Deng, Y.\ 2012, \apj, 745, 164.
\newblock \href{https://doi.org/10.1088/0004-637X/745/2/164}{\path{doi:10.1088/0004-637X/745/2/164}}

\bibitem[Y. Shen et al.(2012)]{2012ApJ...750...12S}
Shen, Y., Liu, Y., \& Su, J.\ 2012, \apj, 750, 12.
\newblock \href{https://doi.org/10.1088/0004-637X/750/1/12}{\path{doi:10.1088/0004-637X/750/1/12}}

\bibitem[Y. Shen et al.(2014a)]{2014ApJ...786..151S}
Shen, Y., Ichimoto, K., Ishii, T.~T., Tian, Z., Zhao, R., \& Shibata, K.\ 2014a, \apj, 786, 151. \newblock \href{https://doi.org/10.1088/0004-637X/786/2/151}{\path{doi:10.1088/0004-637X/786/2/151}}

\bibitem[Y. Shen et al.(2014b)]{2014ApJ...795..130S}
Shen, Y., Liu, Y.~D., Chen, P.~F., \& Ichimoto, K.\ 2014b, \apj, 795, 130.
\newblock \href{https://doi.org/10.1088/0004-637X/795/2/130}{\path{doi:10.1088/0004-637X/795/2/130}}

\bibitem[Y. Shen et al.(2015)]{2015ApJ...814L..17S}
Shen, Y., Liu, Y., Liu, Y.~D., Chen, P.~F., Su, J., Xu, Z., \& Liu, Z.\ 2015, \apjl, 814, L17.
\newblock \href{https://doi.org/10.1088/2041-8205/814/1/L17}{\path{doi:10.1088/2041-8205/814/1/L17}}

\bibitem[Y. Shen et al.(2017)]{2017ApJ...851...67S}
Shen, Y., Liu, Y.~D., Su, J., Qu, Z., \& Tian, Z.\ 2017, \apj, 851, 67.
\newblock \href{https://doi.org/10.3847/1538-4357/aa9a48}{\path{doi:10.3847/1538-4357/aa9a48}}

\bibitem[Y. Shen et al.(2019)]{2019ApJ...873...22S}
Shen, Y., Chen, P.~F., Liu, Y.~D., Shibata, K., Tang, Z., \& Liu, Y.\ 2019a, \apj, 873, 22. \newblock \href{https://doi.org/10.3847/1538-4357/ab01dd}{\path{doi:10.3847/1538-4357/ab01dd}}

\bibitem[Y. Shen et al.(2019)]{2019ApJ...885L..11S}
Shen, Y., Qu, Z., Zhou, C., Duan, Y., Tang, Z., \& Yuan, D.\ 2019b, \apjl, 885, L11.
\newblock \href{https://doi.org/10.3847/2041-8213/ab4cf3}{\path{doi:10.3847/2041-8213/ab4cf3}}

\bibitem[Y. Shen et al.(2019)]{2019ApJ...883..104S}
Shen, Y., Qu, Z., Yuan, D., Chen, H., Duan, Y., Zhou, C., Tang, Z., Huang, J., \& Liu, Y.\ 2019c, \apj, 883, 104.
\newblock \href{https://doi.org/10.3847/1538-4357/ab3a4d}{\path{doi:10.3847/1538-4357/ab3a4d}}

\bibitem[K. Shibata \& Y. Uchida(1985)]{1985PASJ...37...31S}
Shibata, K. \& Uchida, Y.\ 1985, \pasj, 37, 31.

\bibitem[K. Shibata \& Uchida(1986)]{1986SoPh..103..299S}
Shibata, K. \& Uchida, Y.\ 1986, \solphys, 103, 299.
\newblock \href{https://doi.org/10.1007/BF00147831}{\path{doi:10.1007/BF00147831}}

\bibitem[K. Shibata et al.(1992)]{Shibata1992}
Shibata, K., Ishido, Y., Acton, L.~W., et al.\ 1992, \pasj, 44, L173.

\bibitem[K. Shibata et al.(2007)]{Shibata2007}
Shibata, K., Nakamura, T., Matsumoto, T., et al.\ 2007, Science, 318, 1591.
\newblock \href{https://doi.org/10.1126/science.1146708}{\path{doi:10.1126/science.1146708}}

\bibitem[M. Shimojo et al.(2007)]{2007PASJ...59S.745S}
Shimojo, M., Narukage, N., Kano, R., et al.\ 2007, \pasj, 59, S745.
\newblock \href{https://doi.org/10.1093/pasj/59.sp3.S745}{\path{doi:10.1093/pasj/59.sp3.S745}}

\bibitem[A. C. Sterling \& R. L. Moore(2004)]{2004ApJ...602.1024S}
Sterling, A.~C. \& Moore, R.~L.\ 2004, \apj, 602, 1024.
\newblock \href{https://doi.org/10.1086/379763}{\path{doi:10.1086/379763}}

\bibitem[A. C. Sterling et al.(2011)]{2011ApJ...731L...3S}
Sterling, A.~C., Moore, R.~L., \& Freeland, S.~L.\ 2011, \apjl, 731, L3.
\newblock \href{https://doi.org/10.1088/2041-8205/731/1/L3}{\path{doi:10.1088/2041-8205/731/1/L3}}

\bibitem[A. C. Sterling et al.(2015)]{2015Natur.523..437S}
Sterling, A.~C., Moore, R.~L., Falconer, D.~A., \& Adams, M.\ 2015, \nat, 523, 437.
\newblock \href{https://doi.org/10.1038/nature14556}{\path{doi:10.1038/nature14556}}

\bibitem[Y. Su et al.(2014)]{2014ApJ...785L...2S}
Su, Y., G{\"o}m{\"o}ry, P., Veronig, A., Temmer, M., Wang, T., Vanninathan, K., Gan, W., \& Li, Y.\ 2014, \apjl, 785, L2.
\newblock \href{https://doi.org/10.1088/2041-8205/785/1/L2}{\path{doi:10.1088/2041-8205/785/1/L2}}

\bibitem[E. Tandberg-Hanssen(1995)]{1995ASSL..199.....T}
Tandberg-Hanssen, E.\ 1995, The nature of solar prominences, Vol. 199.
\newblock \href{https://doi.org/10.1007/978-94-017-3396-0}{\path{doi:10.1007/978-94-017-3396-0}}

\bibitem[E. Tandberg-Hanssen et al.(1980)]{1980SoPh...65..357T}
Tandberg-Hanssen, E., Martin, S.~F., \& Hansen, R.~T.\ 1980, \solphys, 65, 357.
\newblock \href{https://doi.org/10.1007/BF00152799}{\path{doi:10.1007/BF00152799}}

\bibitem[T{\"o}r{\"o}k \& B. Kliem(2005)]{2005ApJ...630L..97T}
T{\"o}r{\"o}k, T. \& Kliem, B.\ 2005, \apjl, 630, L97.
\newblock \href{https://doi.org/10.1086/462412}{\path{doi:10.1086/462412}}

\bibitem[D. Tripathi et al.(2020)]{2020ApJ...894..128T}
Tripathi, D., Nived, V.~N., Isobe, H., \& Doyle, G.~G.\ 2020, \apj, 894, 128.
\newblock \href{https://doi.org/10.3847/1538-4357/ab8558}{\path{doi:10.3847/1538-4357/ab8558}}

\bibitem[Y. M. Wang(1999)]{1999ApJ...520L..71W}
Wang, Y.-M.\ 1999, \apjl, 520, L71.
\newblock \href{https://doi.org/10.1086/312149}{\path{doi:10.1086/312149}}

\bibitem[D. F. Webb \& A. J. Hundhausen(1987)]{1987SoPh..108..383W}
Webb, D.~F. \& Hundhausen, A.~J.\ 1987, \solphys, 108, 383.
\newblock \href{https://doi.org/10.1007/BF00214170}{\path{doi:10.1007/BF00214170}}

\bibitem[P. F. Wyper et al.(2017)]{2017Natur.544..452W}
Wyper, P.~F., Antiochos, S.~K., \& DeVore, C.~R.\ 2017, \nat, 544, 452.
\newblock \href{https://doi.org/10.1038/nature22050}{\path{doi:10.1038/nature22050}}

\bibitem[T. Yokoyama \& K. Shibata(1995)]{Yokoyama1995}
Yokoyama, T. \& Shibata, K.\ 1995, \nat, 375, 42.
\newblock \href{https://doi.org/10.1038/375042a0}{\path{doi:10.1038/375042a0}}

\bibitem[T. Yokoyama \& K. Shibata(1996)]{Yokoyama1996}
Yokoyama, T. \& Shibata, K.\ 1996, \pasj, 48, 353.
\newblock \href{https://doi.org/10.1093/pasj/48.2.353}{\path{doi:10.1093/pasj/48.2.353}}

\bibitem[P. R. Young \& K. Muglach(2014)]{2014PASJ...66S..12Y}
Young, P.~R. \& Muglach, K.\ 2014, \pasj, 66, S12.
\newblock \href{https://doi.org/10.1093/pasj/psu088}{\path{doi:10.1093/pasj/psu088}}

\bibitem[J. Zhang et al.(2012)]{2012NatCo...3..747Z}
Zhang, J., Cheng, X., \& Ding, M.-D.\ 2012, Nature Communications, 3, 747.
\newblock \href{https://doi.org/10.1038/ncomms1753}{\path{doi:10.1038/ncomms1753}}

\bibitem[J. Zhang et al.(2001)]{2001ApJ...559..452Z}
Zhang, J., Dere, K.~P., Howard, R.~A., Kundu, M.~R., \& White, S.~M.\ 2001, \apj, 559, 452.
\newblock \href{https://doi.org/10.1086/322405}{\path{doi:10.1086/322405}}

\bibitem[Q. M. Zhang et al.(2017)]{2017ApJ...851...47Z}
Zhang, Q.~M., Li, D., \& Ning, Z.~J.\ 2017, \apj, 851, 47.
\newblock \href{https://doi.org/10.3847/1538-4357/aa9898}{\path{doi:10.3847/1538-4357/aa9898}}

\end{thebibliography}

{}



\end{document}